\documentclass[aps,twocolumn,showpacs,floatfix]{revtex4}
\usepackage[latin9]{inputenc}
\usepackage{amsmath}
\usepackage{amssymb}
\usepackage{graphicx}
\usepackage{esint}
\usepackage{epstopdf}
\usepackage{braket}
\usepackage{color}
\usepackage{hyperref}

\makeatletter
\@ifundefined{textcolor}{}
{
 \definecolor{BLACK}{gray}{0}
 \definecolor{WHITE}{gray}{1}
 \definecolor{RED}{rgb}{1,0,0}
 \definecolor{GREEN}{rgb}{0,1,0}
 \definecolor{BLUE}{rgb}{0,0,1}
 \definecolor{CYAN}{cmyk}{1,0,0,0}
 \definecolor{MAGENTA}{cmyk}{0,1,0,0}
 \definecolor{YELLOW}{cmyk}{0,0,1,0}
}

\makeatother

\begin{document}

\title{Entanglement and violation of classical inequalities in the Hawking radiation of flowing
atom condensates}


\author{J. R. M. de Nova, F. Sols, and I. Zapata}
\affiliation{Departamento de Física de Materiales, Universidad Complutense de
Madrid, E-28040 Madrid, Spain}

%
%
%
%
%
%

\begin{abstract}
We consider a sonic black-hole scenario where an atom condensate flows through a subsonic-supersonic interface. We discuss several criteria that reveal the existence of nonclassical correlations resulting from the quantum character of the spontaneous Hawking radiation. We unify previous general work as applied to Hawking radiation analogs. We investigate the measurability of the various indicators and conclude that, within a class of detection schemes, only the violation of quadratic Cauchy-Schwarz inequalities can be discerned. We show numerical results that further support the viability of measuring deep quantum correlations in concrete scenarios.
\end{abstract}


\date{\today}

\maketitle

\section{Introduction}

The detection of spontaneous Hawking radiation (HR) remains a major challenge in modern physics. Originally predicted for cosmological black holes \cite{Hawking1974}, it was soon noticed \cite{Unruh1976,Unruh1981} that the emission of Hawking radiation (HR) is a kinematic effect that can be observed in an ordinary laboratory.  For a quantum fluid traversing
a subsonic-supersonic interface (which amounts to a sonic event horizon), the spontaneous correlated emission of phonons into the subsonic and the supersonic regions
has been predicted  \cite{Garay2000,Leonhardt2003,Leonhardt2003a,Balbinot2008,Carusotto2008,Macher2009,Finazzi2010,Coutant2010}. Sonic event horizons have been realized by accelerating a Bose-Einstein condensate \cite{Lahav2010}. In a similar setup, the self-amplifying stimulated HR resulting from the black-hole (BH) laser effect \cite{Finazzi2010} has been observed \cite{Steinhauer2014}. However, the emission of spontaneous Hawking radiation still remains unobserved.

In the context of quasi-stationary flow scenarios, it has been proposed that a large leaking condensate can provide a suitable subsonic-supersonic interface where HR production could be conveniently observed \cite{Zapata2011,Larre2012}. Realistic protocols to produce such quasi-stationary flow regimes have been investigated recently \cite{deNova2014a}.

It has also been noticed \cite{deNova2014} that the violation of classical Cauchy-Schwarz (CS) inequalities \cite{Loudon1983,Walls2008} by outgoing quasiparticles can provide unambiguous evidence of deep quantum behavior and ultimately of the existence of a spontaneous contribution to HR. The distinction between spontaneous and thermal (stimulated) contributions is a fundamental requirement in the search for spontaneous HR. Proposals based on density-density correlations in real space do not meet that requirement \cite{Balbinot2008,Recati2009}. The mere measurement of phonon \cite{Schutzhold2008} or atom \cite{Zapata2011} intensity spectra would not permit that distinction either.

An alternative scheme to identify deep quantum behavior relies on the detection of entanglement between the various outgoing radiation channels from a sonic horizon \cite{Busch2014,Finazzi2013}. A recent work has addressed the possibility of detecting entanglement through density-density correlations in Fourier space \cite{Steinhauer2015}. Approaches based on entanglement detection have been also proposed in analogous contexts such as inflationary cosmology \cite{Campo2006}, astrophysical black holes \cite{Martin-Martinez2010}, general
relativistic quantum fields \cite{Friis2013}, or other black-hole analogs \cite{Horstmann2010,Horstmann2011}.

Our present work aims at clearly establishing the theoretical relation between CS violation and entanglement in the outgoing quasiparticle modes of analog HR, unifying the existent work of Refs. \cite{Finazzi2013,deNova2014,Busch2014}. We also study their potential measurability in specific detection schemes \cite{Steinhauer2015}. Importantly, by introducing envelope-modulated Fourier transforms we explicitly take into account the role played by the spatial location of the asymptotic regions. This allows us to show that the violation of only quadratic CS inequalities can be measured. In particular, we prove that the complete implementation of the generalized Peres-Horodecki (GPH) criterion \cite{Finazzi2013,Busch2014} or the quartic CS violation \cite{deNova2014} requires the knowledge of parameters that are impossible or very difficult to measure, at least within the class of detection schemes here considered. However, we also show that this does not represent an important limitation in practice, since, under rather broad conditions, all the previous criteria become equivalent.

Although the present work is motivated by the quest for the observation of spontaneous HR, the results here obtained are also of relevance in neighboring fields such as quantum optics \cite{Lopaeva2013,Han2015} and quantum information physics \cite{Clausen2014,Shen2015}, as well as in the broader topic of bosonic condensates \cite{Kheruntsyan2012,Wasak2014}. Atom flow through a sonic black hole may also be viewed as one of the paradigms of atom quantum transport through a barrier \cite{Leboeuf2001}. In general, a subsonic/supersonic interface can behave as a basic element that provides novel functionalities in atom circuits within the emerging field of atomtronics \cite{Seaman2007,Labouvie2015}. For instance, the black-hole configuration of Ref. \cite{deNova2014a} can be used to produce a quasi-stationary supersonic atom current. Another recent numerical work \cite{deNova2015arxiv} has shown that a BH laser configuration is able to reach a regime of continuous emission of solitons, providing a hydrodynamic analog of an optical laser.

This paper is arranged as follows. In Sec. \ref{sec:physicalmodel}, we present the physical model that we use in this work. Section \ref{sec:CSPH} contains a detailed comparison among the existent works on the detection of the spontaneous HR and unifies the criteria under certain conditions. In Section \ref{sec:Experimental}, we discuss the possibility of experimentally verifying the discussed criteria. Finally, in Section \ref{sec:Numerical}, we present numerical results on the detection of the CS violation in typical HR setups.

\section{Physical model}\label{sec:physicalmodel}

We start by considering the stationary flow of a one-dimensional BEC in a regime known as 1D-mean field regime \cite{Menotti2002,Leboeuf2001}. The condensate is described by the stationary GP wave function $\Psi_0(x)=\sqrt{n_0(x)}e^{i\theta(x)}$. We can define the local sound speed and flow velocity as $c(x)=[gn(x)/m]^{1/2}$ and  $v(x)=\hbar \theta'(x)/m$, respectively, with $g$ the coupling constant and $m$ the mass of the atoms. In our convention, $v(x)>0$. A sonic BH configuration is that involving two asymptotically homogeneous regions such that one is subsonic (the upstream region, $x\rightarrow -\infty$), with $v_u<c_u$, while the other one is supersonic (downstream region, $x\rightarrow \infty$), with $v_d>c_d$. The magnitudes $c_{u,d},v_{u,d}$ represent the asymptotic homogeneous values of the sound and flow velocities. Hereafter, we use units such that $\hbar=m=c_u=k_{B}=1$. Throughout this work, we follow the notation of Refs. \cite{Recati2009,Larre2012,Zapata2011}.

\begin{figure*}[tb!]
\includegraphics[width=1\textwidth]{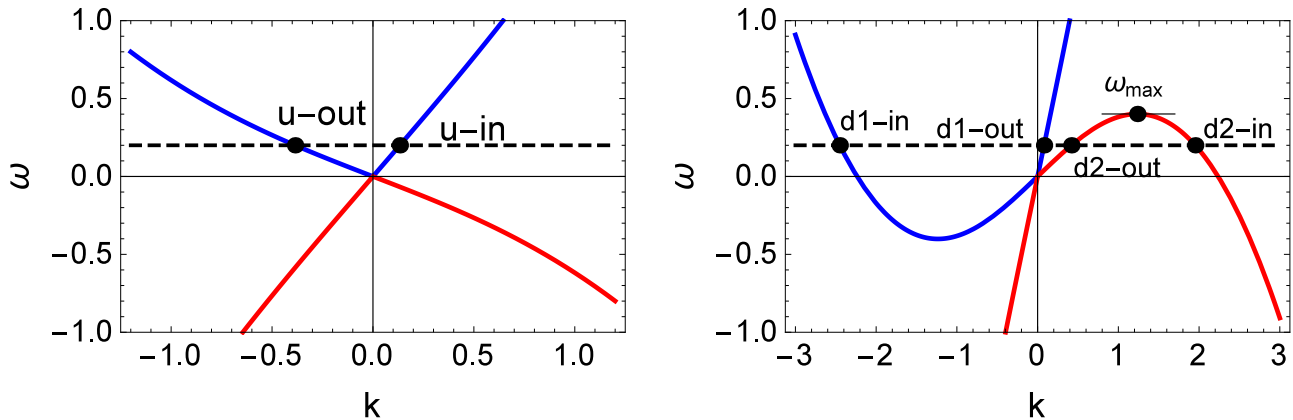}
\caption{Bogoliubov dispersion relation on the subsonic (left, upstream) and supersonic
(right, downstream) sides. The blue (red) branches correspond to positive (negative)
normalization. Here, $d(u)$ denotes downstream(upstream). The Hawking frequency, $\omega_{{\rm max}}$, is the frequency above which
no HR can be generated.}
\label{fig:DispRelation}
\end{figure*}

Within the Bogoliubov-de Gennes (BdG) approximation, the field operator is decomposed as:
\begin{equation}\label{eq:BdGfieldoperator}
\hat{\Psi}(x)=\Psi_0(x)+\hat{\varphi}(x)e^{i\theta(x)}=[\sqrt{n_0(x)}+\hat{\varphi}(x)]e^{i\theta(x)}\, ,
\end{equation}
where we have removed the mean-field condensate phase, $e^{i\theta(x)}$, from the definition of the field fluctuations $\hat{\varphi}(x)$ for computational purposes. The expression of $\hat{\varphi}(x)$ in a BH configuration is given by:
\begin{eqnarray}\label{eq:fieldoperator}
\nonumber \hat{\varphi}(x)& = & \int_{0}^{\infty}d\omega\sum_{I}[u_{I-\rm{in},\omega}(x)\hat{\gamma}_{I-\rm{in}}(\omega)\\
\nonumber &+&v_{I-\rm{in},\omega}^{*}(x)\hat{\gamma}_{I-\rm{in}}^{\dag}(\omega)]\\
\nonumber &+&\int_{0}^{\omega_{{\rm max}}}d\omega[u_{{\rm {d2-\rm{in},\omega}}}(x)\hat{\gamma}_{{\rm {d2-\rm{in}}}}^{\dag}(\omega)\nonumber\\&+&v_{{\rm {d2-\rm{in},\omega}}}^{*}(x)\hat{\gamma}_{{\rm {d2-\rm{in}}}}(\omega)]\, .
\end{eqnarray}
The components of the spinor $z_{a,\omega}(x)\equiv\left[u_{a,\omega}(x),v_{a,\omega}(x)\right]^{\intercal}$ are
solutions to the BdG equations at a given frequency $\omega$.
Here we have adopted the convention of working with positive frequencies at the expense of introducing the negative-normalization channel $d2$. The index $I$ is summed over the conventional, positive-normalization channels ($u$ and $d1$); see Fig. \ref{fig:DispRelation} for the mode notation. In the asymptotically homogeneous regions, the macroscopic wave function has the form
\begin{equation}\label{eq:asymptoticGP}
\Psi_0(x)=\sqrt{n_r(x)}e^{iv_rx+\theta_r},~r=u,d
\end{equation}
where $\theta_r$ is a constant phase. Thus, the scattering states $z_{i,\omega}(x)$ are combinations of plane waves (scattering channels) in the asymptotic regions $u$ and $d$, while $\hat{\gamma}_i(\omega)$ is their corresponding annihilation operator, with the index $i$ taking values $u-,d1-,d2-{\rm in}$. For instance, the scattering state $d2-\rm{in}$ has the asymptotic form
\begin{eqnarray}\label{eq:scatteringstates}
z_{d2-\rm{in},\omega}\left(x\rightarrow-\infty\right)&=&S_{ud2}\left(\omega\right)s_{u-\rm{out},\omega}\left(x\right)\\
z_{d2-\rm{in},\omega}\left(x\rightarrow\infty\right)&=&s_{d2-\rm{in},\omega}\left(x\right)\nonumber\\&+&
S_{d1d2}\left(\omega\right)s_{d1-\rm{out},\omega}\left(x\right)\nonumber\\&+&S_{d2d2}\left(\omega\right)s_{d2-\rm{out},\omega}\left(x\right) \, .\nonumber
\end{eqnarray}
In this expression, $s_{a,\omega}\left(x\right)$ is the spinor (plane) wave function of the corresponding scattering channel $a$:
\begin{eqnarray}\label{eq:scatteringchannels}
s_{a,\omega}\left(x\right) &=& \frac{e^{ik_{a}\left(\omega\right)x}}{\sqrt{2\pi|w_{a}\left(\omega\right)|}}\left[\begin{array}{c}
u_{a}(\omega)\\
v_{a}(\omega)
\end{array}\right] \nonumber\\
\left[\begin{array}{c}
u_{a}(\omega)\\
v_{a}(\omega)
\end{array}\right]&=&N_a\left[\begin{array}{c}
\frac{k_{a}^{2}\left(\omega\right)}{2}+[\omega-v_rk_{a}\left(\omega\right)]\\
\frac{k_{a}^{2}\left(\omega\right)}{2}-[\omega-v_rk_{a}\left(\omega\right)]
\end{array}\right]\\
\nonumber
N_a&=&[2k_{a}^{2}\left(\omega\right)\left|\omega-v_rk_{a}\left(\omega\right)\right|]^{-1/2}\, ,
\end{eqnarray}
and the corresponding wave vector, $k_a(\omega)$ is given implicitly by the dispersion relation
\begin{equation}\label{eq:dispersionrelation}
\Omega^2_a=\left(\omega-v_{r}k_{a}\right)^{2}
=
c_{r}^{2}k_{a}^{2}+k_{a}^{4}/4 \, ,
\end{equation}
where $\Omega_a>0$ is the comoving frequency, defined positive, and the index $r=u,d$ characterizes the asymptotic region corresponding to the scattering channel $a$. The group velocity of the scattering channel $a$ is $w_{a}\left(\omega\right)\equiv\left[dk_{a}\left(\omega\right)/d\omega\right]^{-1}$ and it is included in the definition of the scattering channels in order to properly normalize them to a given quasiparticle flux. The dispersion relation is depicted in Fig. \ref{fig:DispRelation} for the subsonic and supersonic asymptotic regions.

The expressions for the other scattering states are similar to Eq. (\ref{eq:scatteringstates}). They all are characterized by an incoming channel carrying unit flux and several scattering amplitudes (elements of the $S$-matrix) describing the transition to the outgoing modes, which are also assumed to carry unit flux. The ``out'' (advanced) scattering  states are related to the ``in'' (retarded) scattering states through the scattering matrix $S$ for a given value of the frequency $\omega$:
\begin{equation}\label{eq:inoutmodesrelation}
\left[\begin{array}{c}
\hat{\gamma}_{u-\rm{out}}\\
\hat{\gamma}_{d1-\rm{out}}\\
\hat{\gamma}_{d2-\rm{out}}^{\dagger}
\end{array}\right] = \left[\begin{array}{ccc}S_{uu}&S_{ud1}&S_{ud2}\\
S_{d1u}&S_{d1d1}&S_{d1d2}\\
S_{d2u}&S_{d2d1}&S_{d2d2}\end{array}\right]\left[\begin{array}{c}
\hat{\gamma}_{u-\rm{in}}\\
\hat{\gamma}_{d1-\rm{in}}\\
\hat{\gamma}_{d2-\rm{in}}^{\dagger}
\end{array}\right] \, .
\end{equation}
Here, as often in the rest of the paper, the frequency dependence is understood.

We note that Eq. (\ref{eq:inoutmodesrelation}) is a Bogoliubov-type relation linearly combining destruction and creation operators. This is due to the negative normalization of the $d2$ modes, see Fig. \ref{fig:DispRelation}. Thus, the vacuum of the ``in'' modes can be regarded as a squeezed state in the representation of the ``out'' modes. This non-uniqueness of the vacuum is at the root of the Hawking effect itself. In order to detect the intrinsic quantum behavior corresponding to the squeezed character of the zero-point HR, we focus on the correlations between the ``out'' modes.

\section{Cauchy-Schwarz inequalities and entanglement} \label{sec:CSPH}

Several criteria have been proposed in order to distinguish the spontaneous from the stimulated (thermal) HR or the coherent collective modes. First, it was argued in Ref. \cite{deNova2014} that the violation of CS inequalities can be regarded as an unequivocal signature of the presence of zero-point dynamics. Specifically, the following second-order correlation function between two given modes $i,j$ was considered:
\begin{equation}\label{eq:GammaDef}
\Gamma_{ij}\equiv\langle\hat{\gamma}_{i}^{\text{\ensuremath{\dagger}}}
\hat{\gamma}_{j}^{\dagger}\hat{\gamma}_{j}\hat{\gamma}_{i}\rangle>0\, ,
\end{equation}
where $\hat{\gamma}_{i}$ is the annihilation operator of the mode $i$. The expectation value of an operator $\hat{O}$ is taken over the state of the system, described by the density matrix $\hat{\rho}$, $\braket{\hat{O}}=\text{Tr}(\hat{\rho}\hat{O})$. The relation
\begin{equation}\label{eq:CSinequality}
\Gamma_{ij}\leq\sqrt{\Gamma_{ii}\Gamma_{jj}} \, ,
\end{equation}
is a CS-type inequality which is always satisfied in a classical context. The violation of (\ref{eq:CSinequality}) is characterized by the positiveness of the difference
\begin{equation}\label{eq:CSviolation4}
\Theta_{ij}\equiv \Gamma_{ij}-\sqrt{\Gamma_{ii}\Gamma_{jj}}>0 \, .
\end{equation}
We will refer to Eq. (\ref{eq:CSviolation4}) as the {\it quartic} CS violation. We can also define the corresponding first-order correlation functions,
\begin{eqnarray}\label{eq:gdef}
g_{ij}\equiv \braket{\hat{\gamma}_{i}^{\dagger}\hat{\gamma}_{j}},~c_{ij}\equiv \braket{\hat{\gamma}_{i}\hat{\gamma}_{j}} ,
\end{eqnarray}
whose associated {\it quadratic} CS violations, in analogy to Eq. (\ref{eq:CSviolation4}), are given by:
\begin{eqnarray}\label{eq:CSviolation2}
\Delta_{ij}\equiv |c_{ij}|^2-g_{ii}g_{jj}>0 \, .
\end{eqnarray}
The inequality $|g_{ij}|^2\leq g_{ii}g_{jj}$ is always satisfied; see Appendix \ref{app:CSnonseparability}.

Another possible signature of the quantum character of the system is the presence of entanglement between two modes $i,j$. In this work, we follow the convention of defining entanglement as the non-separability of the state of the system. We say that a two-mode state $\hat{\rho}$ is separable when it can be decomposed as:
\begin{equation}\label{eq:separability}
\hat{\rho}=\sum_n p_n\hat{\rho}^{(i)}_n\otimes\hat{\rho}^{(j)}_n
\end{equation}
where $\hat{\rho}^{(i)}_n$ is a state of the Hilbert subspace spanned by the $i$ mode.
The use of the GPH criterion \cite{Horodecki1997,Simon2000} was proposed in Refs. \cite{Finazzi2013,Busch2014} to identify the entanglement between two modes within the context of analog HR emission. The GPH criterion asserts that the state is entangled if $\mathcal{P}_{ij}<0$, where $\mathcal{P}_{ij}$ is the GPH function defined in Eq. (\ref{eq:GPH}). In the particular case of Gaussian states, the reverse implication is also true, i.e., the GPH criterion is also a necessary condition for entanglement \cite{Simon2000}.

Some general remarks about CS violations and entanglement are in order. The violation of a quadratic CS inequality, Eq. (\ref{eq:CSviolation2}), is a sufficient condition for the non-separability of the system \cite{Adamek2013}. In particular, we show in Appendix \ref{app:CSnonseparability} that the quadratic CS violation is a sufficient condition for the fulfillment of the GPH criterion, which in turn is known to imply the presence of entanglement. It can also be shown that the quartic CS violation is independent of the entanglement of the system (see Appendix \ref{app:CSnonseparability}), so separable states can violate quartic CS inequalities and viceversa.

Now we focus on the specific case of analog HR in a BEC. We evaluate the correlation functions of Eqs. (\ref{eq:GammaDef}) and (\ref{eq:gdef}) for the ``out'' modes at given $\omega$:
\begin{eqnarray}\label{eq:CSGamma}
\nonumber\Gamma_{ij}(\omega)&=&\braket{\hat{\gamma}_{i-\rm{out}}^{\text{\ensuremath{\dagger}}}(\omega)\hat{\gamma}_{j-\rm{out}}^{\dagger}(\omega)\hat{\gamma}_{j-\rm{out}}(\omega)\hat{\gamma}_{i-\rm{out}}(\omega)}\\
g_{ij}(\omega)&\equiv&\braket{\hat{\gamma}_{i-\rm{out}}^{\dagger}(\omega)\hat{\gamma}_{j-\rm{out}}(\omega)}\\
\nonumber c_{ij}(\omega)&\equiv&\braket{\hat{\gamma}_{i-\rm{out}}(\omega)\hat{\gamma}_{j-\rm{out}}(\omega)} \, .
\end{eqnarray}
Hereafter, $i,j=u,d1,d2$. The associated quartic and quadratic CS violations are characterized by the positivity of $\Theta_{ij}(\omega)$ and $\Delta_{ij}(\omega)$, respectively, as defined in Eqs. (\ref{eq:CSviolation4}),(\ref{eq:CSviolation2}). In a similar fashion, we study the GPH function $\mathcal{P}_{ij}(\omega)$ for the $i,j$  ``out'' modes at the same frequency $\omega$. For simplicity, in this section we will obviate the Dirac delta factors appearing in all equal-frequency correlation functions [see Eq. (\ref{eq:open-bc}) for a complete expression].

One may wonder to what extent the previous criteria differ when applied to spontaneous HR. We devote the rest of this section to prove that, when the state of the system $\hat{\rho}$ is Gaussian and incoherent in the ``in'' modes, the quartic and quadratic CS violations become equivalent to the GPH criterion. Specifically, the requirement of incoherent incoming modes can be expressed as:
\begin{eqnarray} \label{eq:open-bc}
\nonumber \braket{\hat{\gamma}_{i-\rm{in}}(\omega)\hat{\gamma}_{j-\rm{in}}(\omega')}&=&0\\
\braket{\hat{\gamma}_{i-\rm{in}}^{\dagger}(\omega)\hat{\gamma}_{j-\rm{in}}(\omega')}
&=&
n_i(\omega)\delta_{ij}\delta(\omega-\omega') \neq0\, .
\end{eqnarray}
These correlation functions fully characterize the state provided it is Gaussian. It is shown in Appendix \ref{app:parametrization} that, for that class of states, we only need $7$ parameters for the computation of the CS violations and the GPH function. The case of Gaussian states is important because, within the BdG approximation, the dynamics is Gaussian.

\begin{table}[!tb]
\begin{tabular}[c]{|c|c|c|c|c|}
\hline
Gaussian & incoherent & $\text{CS4}$ & $\text{CS2}$ & GPH\\
\hline
yes & yes &~$\bullet$ $\Leftrightarrow$ GPH &~ $\bullet$ $\Leftrightarrow$ GPH &~ $\bullet$ $\Leftrightarrow$ NS\\
\hline
yes & no & ~$\bullet$ indep.  NS &~ $\bullet$ $\Rightarrow$ GPH &~ $\bullet$ $\Leftrightarrow$ NS\\
\hline
no & yes &~$\bullet$ indep.  NS &~ $\bullet$ $\Leftrightarrow$ GPH &~ $\bullet$ $\Rightarrow$ NS\\
\hline
no & no & ~$\bullet$ indep.  NS &~ $\bullet$ $\Rightarrow$ GPH &~ $\bullet$ $\Rightarrow$ NS\\
\hline
\end{tabular}
\caption{Logical relations between the different criteria studied throughout this work in the various physical cases. The three rightmost entries in the upper row indicate the different criteria for the identification of quantum behavior:
$\text{CS2}$, $\text{CS4}$ stand for quadratic and quartic CS violations, respectively, while GPH stands for generalized Peres-Horodecki criterion; NS means non-separability.
The two leftmost columns define
the various physical cases here considered. By ``incoherent'' state, we understand a density matrix which is diagonal in the representation of retarded quasiparticle scattering modes, each characterized by a single incoming channel; see Eq. (\ref{eq:open-bc}). The symbol $\bullet$ stands for the uppermost entry in the corresponding column.
The abbreviation ``indep.'' means that, in the three lower cases, the quartic CS violation is independent of the non-separability of the system.}
\label{table}
\end{table}

For convenience, we define the complex vector
\begin{eqnarray}\label{eq:defcorrs}
\alpha_i(\omega)&\equiv&\left[\begin{array}{c}
S_{iu}(\omega)\sqrt{n_u(\omega)}\\
S_{id1}(\omega)\sqrt{n_{d1}(\omega)}\\
S_{id2}(\omega)\sqrt{n_{d2}(\omega)+1}
\end{array}\right]\, .
\end{eqnarray}
From the previous definitions, it is easy to show that the only non-zero quadratic correlations for ``out'' modes are:
\begin{eqnarray}\label{eq:quadraticorrs}
\nonumber g_{II'}&=& \alpha_{I}^{\dagger}\cdot\alpha_{I'} \\
\nonumber g_{d2d2}&=&|\alpha_{d2}|^{2}-1\\
c_{Id2}&=&\alpha_{d2}^{\dagger}\cdot\alpha_{I}\, ,
\end{eqnarray}
where the index $I$ stands for a normal, positive-normalization ($u$ or $d1$) mode.
First, we study the correlation between a normal and an anomalous (negative-normalization) mode.
The case $I=u$ corresponds to the proper Hawking effect \cite{Recati2009,Zapata2011} and the case $I=d1$ corresponds to the bosonic equivalent of the Andreev reflection \cite{Zapata2009a}.

As we are working with Gaussian states, we can apply Wick's theorem to compute the quartic correlations as a function of the quadratic correlations
\begin{eqnarray}\label{eq:CSWick}
\nonumber \Gamma_{II} & = & 2g^2_{II}=2|\alpha_{I}|^{4}\\
\nonumber \Gamma_{d2d2} & = & 2g^2_{d2d2}=2(|\alpha_{d2}|^{2}-1)^{2}\\
\nonumber \Gamma_{Id2} & = & |c_{Id2}|^2+g_{II}g_{d2d2}\\
&=&|\alpha_{d2}^{\dagger}\cdot\alpha_{I}|^{2}+|\alpha_{I}|^{2}(|\alpha_{d2}|^{2}-1)\, .
\end{eqnarray}
Therefore, the condition for the quartic CS violation $\Theta_{Id2}(\omega)>0$ reduces to the simpler quadratic CS violation
\begin{equation}\label{eq:CSnonseparable}
\Theta_{Id2}(\omega)=\Delta_{Id2}(\omega)>0
\end{equation}
which, using Eqs. (\ref{eq:quadraticorrs}) and (\ref{eq:CSWick}), can be rewritten as:
\begin{equation}\label{eq:CSnonseparablealphas}
|\alpha_{d2}^{\dagger}\cdot\alpha_{I}|^{2}>|\alpha_{I}|^{2}(|\alpha_{d2}|^{2}-1)
\end{equation}
The $-1$ within the second bracket results from the anomalous character of the $d2$ mode and is responsible for making the violation of the CS inequality possible. Thus, we have proven that the quartic and the quadratic violations are equivalent for $Id2$, normal-anomalous correlations.

Now we turn our attention to the GPH criterion. From Eq. (\ref{eq:GPHcomplete}) we compute the GPH function for the pair of modes $Id2$:
\begin{equation}\label{eq:GPHanomalous1}
\mathcal{P}_{Id2}=(g_{II}g_{d2d2}-|c_{Id2}|^2)[(g_{II}+1)(g_{d2d2}+1)-|c_{Id2}|^2] \, .
\end{equation}

As the second (square) bracket in the r.h.s. of Eq. (\ref{eq:GPHanomalous1}) is always positive [see Eq. (\ref{eq:CScijpos})], we conclude that $\mathcal{P}_{Id2}<0$ if and only if Eq. (\ref{eq:CSnonseparable}) is satisfied. As the state is Gaussian, the GPH criterion is equivalent to the entanglement of the state of the system $\hat{\rho}$. We conclude that the quadratic and quartic CS violations are equivalent to entanglement. We note that the equivalence between condition (\ref{eq:CSnonseparable}) and entanglement was already pointed out in Ref. \cite{Busch2014} but the connection with the CS violation criterion of Ref. \cite{deNova2014} was not made explicit. A similar result appeared in Ref. \cite{Busch2014a} on the equivalence between non-separability and CS violation when studying the correlation function at two different times in fluids of light.

In regard to the correlation between the two normal modes, we obtain, following similar arguments,
that the quartic CS inequality (\ref{eq:CSinequality}) reduces to:
\begin{equation}\label{eq:CSnormal}
|g_{ud1}|^2\leq g_{uu}g_{d1d1} \, ,
\end{equation}
which can never be violated, as previously explained [see Eq. (\ref{eq:CSviolation2}) and accompanying discussion]. In particular, the inequality (\ref{eq:CSnormal}) can be rewritten as:
\begin{equation}\label{eq:CSnonviolation}
|\alpha_{u}^{\dagger}\cdot\alpha_{d1}|^2\leq |\alpha_{u}|^2|\alpha_{d1}|^2 \, ,
\end{equation}
which is always satisfied for two complex vectors. Thus, there is no CS violation in the correlation between normal modes.

When considering the entanglement between the two normal modes, we have $|g_{ud1}|>|c_{ud1}|=0$ and thus there is no entanglement [see Eq. (\ref{eq:GPHcomplete}) and the ensuing discussion].

From the previous discussion we conclude that, when considering Gaussian and incoming incoherent states [i.e., states satisfying Eq. (\ref{eq:open-bc})], all the here considered criteria for characterizing the quantum character of the system become equivalent. This result is important because it unifies the work of Refs. \cite{deNova2014,Finazzi2013,Busch2014}.

A most important particular case is that involving a thermal distribution of incoming quasiparticles that have thermalized in the comoving reference frame \cite{Macher2009a,Busch2014} so that  their occupation factor is:
\begin{equation}\label{eq:thermaldistribution}
n_{i}(\omega)=\frac{1}{\exp(\Omega_{i}(\omega)/T)-1}
\end{equation}
with $\Omega_{i}(\omega)$ the commoving frequency of the mode $i$-in at laboratory frequency $\omega$, as given by Eq. (\ref{eq:dispersionrelation}). This state satisfies the mentioned conditions, so there is no difference between using the GPH criterion or the CS violation to characterize the quantum behavior of the system.

Finally, we discuss the differences that appear when removing some restrictions. If the state is Gaussian but not incoherent in the incoming channels, the GPH criterion is still equivalent to the non-separability of the system. On the other hand, in the same case (Gaussian and not incoherent), the quadratic CS violation is no longer equivalent to the GPH criterion; rather, it is only a sufficient condition for it. As a consequence, the GPH criterion is a more powerful criterion than the quadratic CS violation to detect nonseparability (it can identify nonseparability in cases where the quadratic CS violation would fail).

However, when relaxing the requirement of incoherence, the quartic CS violation becomes independent of the GPH criterion (see Appendix \ref{app:CSnonseparability}).
In this context, we wish to note, using quantum optics terminology, that the presence of quartic CS violation or entanglement requires that the system is described by a non-classical Glauber-Sudarshan function, i.e., a Glauber-Sudarshan function that takes negative values.

On the other hand, if the state of the system is not Gaussian but is incoherent in the incoming channels, the quartic CS violation is also independent of the entanglement. The quadratic CS violation Eq. (\ref{eq:CSnonseparable}) and the GPH criterion are still equivalent between them, but no longer equivalent to the entanglement of the system.

For a general state which does not satisfy any of the previous conditions, the quadratic CS violation is only a sufficient condition for the GPH criterion, which in turn is a sufficient condition for the presence of entanglement. We summarize all these logical relations in Table \ref{table}.

Remarkably, the quadratic CS violation Eq. (\ref{eq:CSnonseparable}) reveals at the same time two different aspects of quantum behavior: the violation of a classical inequality and the entanglement of the system.

\section{On the experimental detection} \label{sec:Experimental}

In this section, we analyze possible detection schemes of the criteria discussed in the previous section. A particular type of CS violation between two colliding condensates was measured using time-of-flight detection \cite{Kheruntsyan2012}. The possible detection of the quartic CS violation in a TOF experiment for analog HR was discussed in Ref. \cite{deNova2014}. A recent work has studied the measurement of quartic CS violations \cite{Boiron2015} using phonon evaporation, following a related work on the dynamical Casimir effect \cite{Jaskula2012}. On the other hand, the detection of entanglement using the GPH criterion was analyzed in Ref. \cite{Finazzi2013} using density fluctuations or the optomechanical detection of phonons. Recently, it has been proposed that the entanglement of Hawking radiation in a BEC can be detected experimentally by measuring the density-density correlations through \textit{in situ} imaging \cite{Steinhauer2015}. All these setups have in common that they involve collective atom flow and thus can be viewed as elementary components of a larger atomtronic circuit.

For illustrative purposes, we focus on the comparison between the detection of the CS violation and the GPH criterion for the particular scenario of density-density correlations, but the analysis which we present in this work can be generalized to other schemes. The role of the spatial density correlation function in analog models has been extensively studied in several works \cite{Balbinot2008,Recati2009,Larre2012}. In the BEC context, the measurement of the spatial density correlations has been experimentally used to characterized the BH Laser \cite{Steinhauer2014}. The density-density correlation function also plays an important role in the polariton analog \cite{Gerace2012,Nguyen2015}.

We start by considering the density in the same BdG approximation of Sec. \ref{sec:physicalmodel} and expand up to first order in the fluctuations of the field operator:
\begin{equation}\label{eq:BdGdensity}
\hat{n}(x)\simeq n_0(x)+\sqrt{n_0(x)}\hat{\phi}(x),~\hat{\phi}(x)=\hat{\varphi}(x)+\hat{\varphi}^{\dagger}(x)
\end{equation}
The expression of $\hat{\phi}(x)$ in terms of the BdG scattering states is formally similar to that of $\hat{\varphi}(x)$, Eq. (\ref{eq:fieldoperator}):

\begin{eqnarray}\label{eq:densityoperator}
\hat{\phi}(x)& = & \int_{0}^{\infty}d\omega\sum_{I} r_{I-\rm{in},\omega}(x)\hat{\gamma}_{I-\rm{in}}(\omega)\\
\nonumber &+&\int_{0}^{\omega_{{\rm max}}}d\omega~r_{{\rm {d2-\rm{in},\omega}}}(x)\hat{\gamma}_{{\rm {d2-\rm{in}}}}^{\dag}(\omega)
+ {\rm H.c.} \, .
\end{eqnarray}
with $r_{a,\omega}(x)=u_{a,\omega}(x)+v_{a,\omega}(x)$.
Importantly, we note that the density in the BdG approximation is linear in the destruction operators, rather than quadratic. Thus, we can extract the quadratic correlations of Sec. \ref{sec:CSPH} by measuring density-density correlations \cite{Finazzi2013,Steinhauer2015}. For that purpose, we restrict ourselves to asymptotic (subsonic and supersonic) regions, where the GP wave function adopts the form of Eq. (\ref{eq:asymptoticGP}). We refer to the region between the two asymptotic regions as the scattering region, where the sonic black hole is placed; i.e., within the scattering region the flow velocity crosses the sound velocity at least once. In the following, we assume that the scattering region is placed near $x=0$ and its size is much smaller than the size of the asymptotic, homogeneous regions.

By taking the Fourier transform of the density in the asymptotic regions at $k\neq 0$, we can get rid of the condensate signal and extract the phonon signal. For definiteness, we focus on the correlations $u-d2$ but the procedure for the other cases is similar. We define:
\begin{equation}\label{eq:Fourierdensity}
\hat{n}_{r}\left(k\right)\equiv\int\mathrm{d}x\, \hat{n}(x)f_{r}^{*}\left(x\right)e^{-ikx}\, ,
\end{equation}
where $f_{r}\left(x\right)$ are normalized functions ($\int\mathrm{d}x|f_{r}(x)|^{2}=1$) localized in the corresponding asymptotic homogeneous regions, sufficiently far from the scattering region.
They represent the envelope of the Fourier transform, which has to be introduced to explicitly take into account the fact that the asymptotic subsonic and supersonic regions are placed in different spatial regions in a realistic situation. We choose them such that
\begin{equation}\label{eq:envelope}
f_{r}(x)=\frac{1}{\sqrt{L_{r}}}f\left(\frac{x-x_{r}}{L_{r}}\right) \, ,
\end{equation}
with $f$ a symmetric and real dimensionless function, $x_{r}$ the point
where the envelope is centered, and $L_{r}$ the size of the wave packet, which is taken sufficiently large for the
Fourier transform of the envelope function,
\begin{equation}\label{eq:Fouriertransformenvelope}
f_{r}(k)=\frac{1}{\sqrt{2\pi}}\int\mathrm{d}x~e^{-ikx}f_{r}\left(x\right)\, ,
\end{equation}
to be sufficiently well peaked at zero momentum. Since the subsonic (supersonic) region is placed at the left (right) of the BH, we have $x_u<0$ ($x_d>0$).

\begin{figure*}[!tb]
\begin{tabular}{@{}cc@{}}
    \includegraphics[width=\columnwidth]{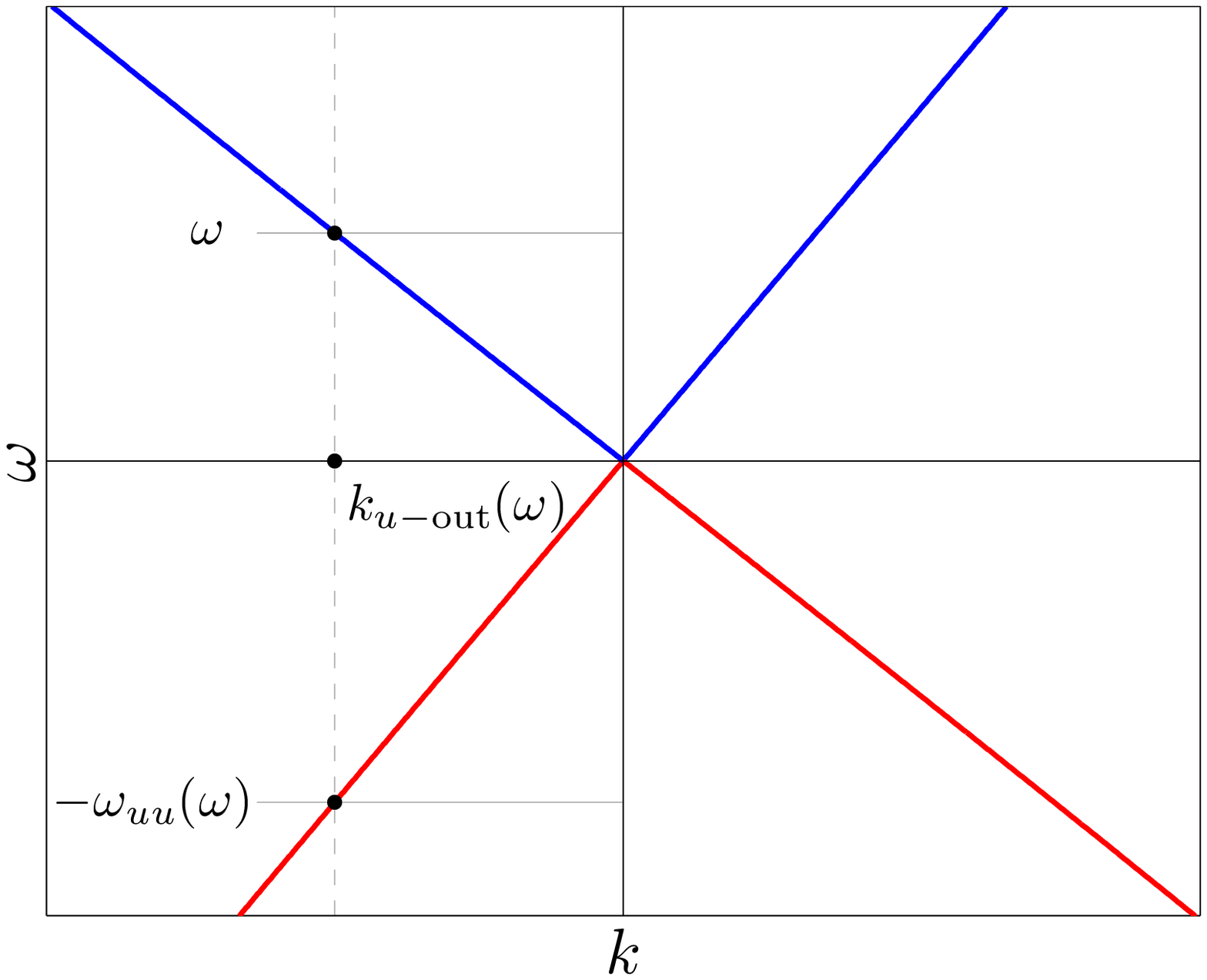} &
    \includegraphics[width=0.963\columnwidth]{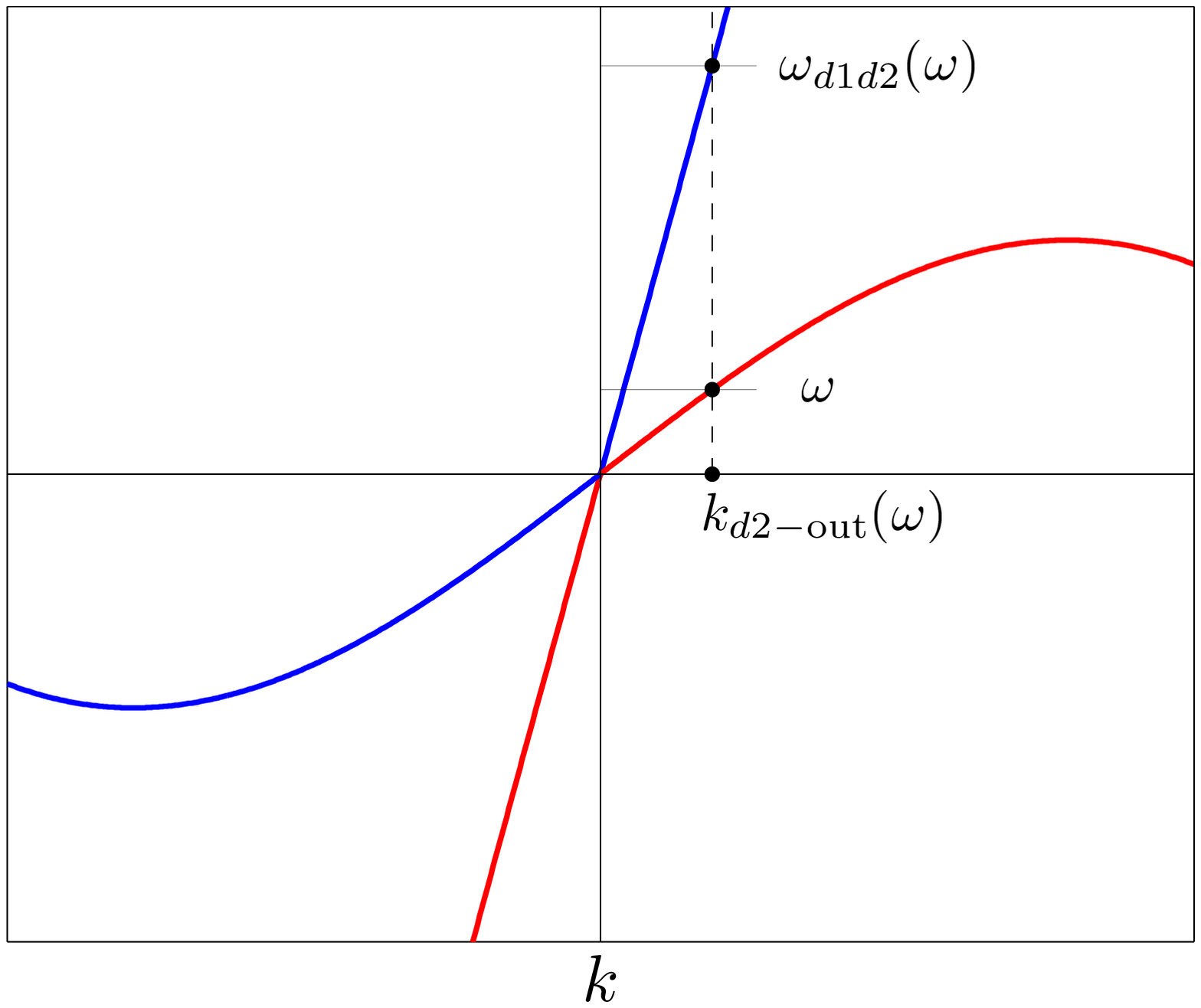} \\
\end{tabular}
\caption{Graphical obtention of $\omega_{d1d2}(\omega)$ and $-\omega_{uu}(\omega)$.
See Fig. \ref{fig:DispRelation} for mode notation. }
\label{fig:DepletionOmegas}
\end{figure*}

Taking into account all the previous considerations, we obtain:
\begin{widetext}
\begin{eqnarray}\label{eq:density}
\nonumber\hat{n}_{u}\left(k_{u-\rm{out}}(\omega)\right) & \simeq & \sqrt{n_u}\int\mathrm{d}\omega'~f^{*}_u(k_{u-\rm{out}}(\omega')-k_{u-\rm{out}}(\omega))\frac{r_{u-\rm{out}}(\omega')}
{\left|w_{u-\rm{out}}\left(\omega'\right)\right|^{1/2}}
\hat{\gamma}_{u-\rm{out}}(\omega')\\
 \nonumber& + & f^{*}_u(-k_{u-\rm{in}}(\omega')-k_{u-\rm{out}}(\omega))
\frac{r_{u-\rm{in}}(\omega')}
{|w_{u-\rm{in}}\left(\omega'\right)|^{1/2}}
\hat{\gamma}_{u-\rm{in}}^{\dag}(\omega')\\
\nonumber\hat{n}_{d}\left(-k_{d2-\rm{out}}(\omega)\right) & \simeq & \sqrt{n_d}\int_{0}^{\omega_{\rm max}}\mathrm{d}\omega'~f^{*}_d(k_{d2-\rm{out}}(\omega)-k_{d2-\rm{out}}(\omega'))
\frac{r_{d2-\rm{out}}(\omega')}
{|w_{d2-\rm{out}}\left(\omega'\right)|^{1/2}}
\hat{\gamma}_{d2-\rm{out}}(\omega')\\
 & + & \int~\mathrm{d}\omega'~f^{*}_d(k_{d2-\rm{out}}(\omega)-k_{d1-\rm{out}}(\omega'))
 \frac{r_{d1-\rm{out}}(\omega')}
 {|w_{d1-\rm{out}}\left(\omega'\right)|^{1/2}}
 \hat{\gamma}_{d1-\rm{out}}^{\dag}(\omega'),
\end{eqnarray}
\end{widetext}
where $r_a(\omega)$ is [see Eq. (\ref{eq:scatteringchannels})]
\begin{equation} \label{r-a}
r_{a}(\omega)=u_{a}(\omega)+v_{a}(\omega)=\sqrt{\frac{k^2_a}{2\Omega_a(\omega)}},
\end{equation}
and we have used $|\omega-v_rk_a(\omega)|=\Omega_a(\omega)=\sqrt{c_r^2k^2+\frac{k^4}{4}}$. Now, we can connect the correlations studied in Sec. \ref{sec:CSPH} with the density correlations by taking into account that $\hat{n}^{\dagger}_{r}(k)=\hat{n}_{r}(-k)$. We obtain:
\begin{widetext}
\begin{eqnarray}\label{eq:measuredcorrelations}
\nonumber G_{uu}(\omega)\equiv\braket{\hat{n}_{u}\left(-k_{u-\rm{out}}\left(\omega\right)\right)\hat{n}_{u}\left(k_{u-\rm{out}}\left(\omega\right)\right)}&=& n_ur^2_{u-\rm{out}}(\omega)[g_{uu}(\omega)+n_{u}(\omega_{uu}(\omega))+1]\\
G_{d2d2}(\omega)\equiv\braket{\hat{n}_{d}\left(k_{d2-\rm{out}}(\omega)\right)\hat{n}_{d}\left(-k_{d2-\rm{out}}(\omega)\right)} & = & n_dr^2_{d2-\rm{out}}(\omega)[g_{d2d2}(\omega)+g_{d1d1}(\omega_{d1d2}(\omega))+1] \\
\nonumber G_{ud2}(\omega)\equiv\braket{\hat{n}_{d}\left(-k_{d2-\rm{out}}(\omega)\right)\hat{n}_{u}\left(k_{u-\rm{out}}\left(\omega\right)\right)} & = & F(\omega)\sqrt{n_un_d}r_{u-\rm{out}}r_{d2-\rm{out}}c_{ud2}(\omega) \, ,
\end{eqnarray}
\end{widetext}
with $\omega_{d1d2}(\omega)=\omega_{d1}(k_{d2-\rm{out}}(\omega))$ and
$\omega_{uu}(\omega)=\omega_{u}(-k_{u-\rm{out}}(\omega))$, $\omega_{i}(k)$ being
the dispersion relation of the mode $i$. Their corresponding values are obtained graphically in Fig. \ref{fig:DepletionOmegas}.
On the other hand, the overlap function between the subsonic and supersonic regions, $F(\omega)$, is given by:
\begin{eqnarray}\label{eq:overlap}
&&F(\omega) = \int\mathrm{d}k~f_{u}^{*}\left(\frac{k}{\zeta(\omega)}\right)f_{d}^{*}\left(k\zeta(\omega)\right)
\\ \nonumber
&=& \frac{1}{\sqrt{L_uL_d}}
\int\mathrm{d}x\, f^{*}\left(\frac{\zeta\left(\omega\right)x+x_{u}}{L_{u}}\right)f^{*}\left(\frac{\zeta\left(\omega\right)^{-1}x-x_{d}}{L_{d}}\right)\nonumber \, ,
\end{eqnarray}
with
\begin{equation}
\zeta(\omega) \equiv\left|\frac{w_{u-\rm{out}}(\omega)}{w_{d2-\rm{out}}(\omega)}\right|^{1/2} \, .
\end{equation}

The previous integral can be interpreted as
the scalar product of two normalized functions, which satisfies $F(\omega)\leq1$.
It is easy to prove that $F(\omega)=1$ when
\begin{equation}\label{eq:densitylines}
\zeta(\omega)=\sqrt{\frac{L_{u}}{L_{d}}}=\sqrt{\left|\frac{x_{u}}{x_{d}}\right|},
\end{equation}
This condition has been interpreted as the condition for the wave-packets to have equal width in frequency space \cite{deNova2014}. However, resulting from the well-known properties of the spatial density correlations, Eq. (\ref{eq:densitylines}) can be also interpreted as expressing that the envelopes in the subsonic and supersonic regions have to be placed along the correlation lines that maximize the spatial density correlation function, see Refs. \cite{Balbinot2008,Recati2009,Larre2012}.

In the following, we suppose that condition (\ref{eq:densitylines}) is satisfied and $F(\omega)=1$. It is worth noting that the correlation functions $G_{ij}(\omega)$ can never violate the CS inequality by themselves since $G_{ud2}<\sqrt{G_{uu}G_{d2d2}}$. Thus, we try to relate these functions, which can be measured, to the phonon correlations considered in the previous section. For that purpose, we normalize the density-density correlations of Eq. (\ref{eq:measuredcorrelations}):
\begin{equation}\label{eq:normalizedcor}
\mathcal{G}_{ij}(\omega) \equiv \frac{G_{ij}(\omega)}{\sqrt{n_in_j}r_{i-\rm{out}}(\omega)r_{j-\rm{out}}} \, ,
\end{equation}
where $n_{i,j}$ must be interpreted as $n_{i,j}=n_d$ when the indices $i,j$ take values $d1,d2$.
From Eq. (\ref{eq:measuredcorrelations}), we see that $\mathcal{G}_{ud2}(\omega)=c_{ud2}(\omega)$. For the extraction of the other correlations, in a similar way to Ref. \cite{Steinhauer2015}, we define the magnitudes:
\begin{eqnarray}\label{eq:measuredcorrelations-1}
\nonumber \tilde{g}_{uu}(\omega)&\equiv&g_{uu}(\omega)+n_{u}(\omega_{uu}(\omega))=\mathcal{G}_{uu}(\omega)-1\\
\nonumber \tilde{g}_{d2d2}(\omega)&\equiv&g_{d2d2}(\omega)+g_{d1d1}(\omega_{d1d2}(\omega))=\mathcal{G}_{d2d2}(\omega)-1 \\
\end{eqnarray}
The $-1$ appearing in the definitions of $\tilde{g}_{uu}(\omega),\tilde{g}_{d2d2}(\omega)$ reflects the subtraction of the atomic depletion contribution.

We see that $\tilde{g}_{uu},\tilde{g}_{d2d2}$ are over-estimations of the correlation functions $g_{uu},g_{d2d2}$, since $n_{u},g_{d1d1}$ are always positive. From this, we can define
\begin{equation}\label{eq:Delta-ud2}
\tilde{\Delta}_{ud2}\equiv|c_{ud2}|^2-\tilde{g}_{uu}\tilde{g}_{d2d2} \, , \\
\end{equation}
noting that $\Delta_{ud2}>\tilde{\Delta}_{ud2}$. Thus, measuring $\tilde{\Delta}_{ud2}>0$ amounts to experimentally observing the quadratic CS violation. We note that, for these calculations, the sole assumption has been made that the state of the system is stationary.

The upshot of this discussion is that we can observe the violation of a quadratic CS inequality through the measurement of the function $\tilde{\Delta}_{ud2}$. Similar claims can be made about the measurement of the quadratic CS inequality involving the $d1,d2$ modes. We can repeat the same strategy as before to obtain the correlation functions $\tilde{g}_{d1d1},c_{d1d2}$. However, it is not possible to obtain the correlation functions $c_{uu},c_{d1d1},c_{d2d2},g_{ud2},g_{d1d2}$ from this scheme. This is due to the vanishing overlap integrals that appear when trying to obtain the corresponding correlations. For instance, if we compute $\braket{\hat{n}_{u}\left(k_{u-\rm{out}}\left(\omega\right)\right)\hat{n}_{u}\left(k_{u-\rm{out}}\left(\omega\right)\right)}$, we face an overlap integral of the type
\begin{equation}\label{eq:overlap}
\int\mathrm{d}k~f_{u}^{*}\left(k\right)f_{u}^{*}\left(k\right)=\int\mathrm{d}x\, f^*_u\left(x\right)f^*_u\left(-x\right)=0,
\end{equation}
because the $f_{u}$ function is, by construction, well localized in the subsonic region, far from the scattering region around $x=0$, see Eq. (\ref{eq:Fourierdensity}) and the discussion below. Thus, we cannot obtain $c_{uu}$ by this method. A similar reasoning applies to other correlations.

It is important to remark on the crucial role played by the spatial location of the asymptotic regions. If we did not introduce the envelopes in Eq. (\ref{eq:Fourierdensity}) and rather did take the Fourier transforms as ideally infinite, we would obtain:
\begin{widetext}
\begin{eqnarray}\label{eq:densitynaive} 
\hat{n}_{u}\left(k_{u-\rm{out}}(\omega)\right) & \simeq & \sqrt{n_u}\left(\frac{r_{u-\rm{out}}(\omega)}{\left|w_{u-\rm{out}}\left(\omega\right)\right|^{1/2}}\hat{\gamma}_{u-\rm{out}}(\omega)+
\frac{r_{u-\rm{in}}(\omega_{uu}(\omega))}
{|w_{u-\rm{in}}\left(\omega_{uu}(\omega)\right)|^{1/2}}
\hat{\gamma}_{u-\rm{in}}^{\dag}(\omega_{uu}(\omega))\right)\\
\nonumber\hat{n}_{d}\left(-k_{d2-\rm{out}}(\omega)\right) & \simeq & \sqrt{n_d}\left(
\frac{r_{d2-\rm{out}}(\omega)}
{|w_{d2-\rm{out}}\left(\omega\right)|^{1/2}}
\hat{\gamma}_{d2-\rm{out}}(\omega)+\frac{r_{d1-\rm{out}}(\omega_{d1d2}(\omega))}
{|w_{d1-\rm{out}}\left(\omega_{d1d2}(\omega)\right)|^{1/2}}
\hat{\gamma}_{d1-\rm{out}}^{\dag}(\omega_{d1d2}(\omega))\right),
\end{eqnarray}

We can infer that such an approach would lead to contradictory results. For instance, the commutator
\begin{equation}\label{eq:contradiction}
[\hat{n}_{u}\left(k_{u-\rm{out}}(\omega)\right),\hat{n}_{u}\left(-k_{u-\rm{in}}(\omega)\right)]=n_u\frac{r_{u-\rm{out}}(\omega)}{\left|w_{u-\rm{out}}\left(\omega\right)\right|^{1/2}}
\frac{r_{u-\rm{in}}(\omega)}
{|w_{u-\rm{in}}\left(\omega\right)|^{1/2}}S_{uu}(\omega)\neq 0
\end{equation}
would be non-vanishing, in contradiction with the well-known fact that it has to be zero since $[\hat{n}(x),\hat{n}(x')]=0$ for every $x,x'$. By way of contrast, when using the full expressions of Eq. (\ref{eq:density}) we arrive at the correct result
\begin{equation}\label{eq:consistent}
[\hat{n}_{u}\left(k_{u-\rm{out}}(\omega)\right),\hat{n}_{u}\left(-k_{u-\rm{in}}(\omega)\right)]=n_u\frac{r_{u-\rm{out}}(\omega)}{\left|w_{u-\rm{out}}\left(\omega\right)\right|^{1/2}}
\frac{r_{u-\rm{in}}(\omega)}
{|w_{u-\rm{in}}\left(\omega\right)|^{1/2}}S_{uu}(\omega)\int\mathrm{d}\omega'~f_{u}^{*}\left(-\frac{\omega'-\omega}{|w_{u-\rm{in}}\left(\omega\right)|}\right)f_{u}^{*}\left(-\frac{\omega'-\omega}{|w_{u-\rm{out}}\left(\omega\right)|}\right)
\end{equation}
\end{widetext}
The integral of the r.h.s, after returning to real space, is similar to that considered in Eq. (\ref{eq:overlap}) and gives zero.

We arrive at the same conclusion when considering the density-density correlations in real space. As noted in Ref. \cite{Recati2009}, only the terms with a stationary phase should be kept when computing $\braket{\hat{n}(x)\hat{n}(x')}$, which yields the condition (\ref{eq:densitylines}). In particular, only the ``proper'' correlations between out-out modes can be extracted $c_{ud2},c_{d1d2},g_{ud1}$. The other correlations $c_{uu},c_{d1d1},c_{d2d2},g_{ud2},g_{d1d2},c_{ud1}$, even when they are non-zero, cannot be obtained because the associated exponential terms do not present a stationary phase when integrating over frequencies.

We conclude that an important consequence of taking into account the spatial location of the subsonic and supersonic regions is that only a limited number of correlations can be experimentally observed. In particular, this implies that we cannot measure all the correlation functions appearing in the GPH criterion (\ref{eq:GPH}) by this procedure. Moreover, and for the same reason, if the state is Gaussian, we cannot obtain all the correlation functions appearing in the quartic CS violation, see Eq. (\ref{eq:CSviolation4Wick}).

The results of the present section show that, within this kind of detection schemes, we can only aim at observing a quadratic CS violation. One can expect this
limitation to be not exclusive of the specific procedure here considered. The reason is that any realistic attempt to obtain the correlation functions between phonons must necessarily take into account the spatial location of the asymptotic regions, which implies that a similar reasoning does apply.

Nevertheless, the previous considerations do not pose a problem for the detection of the quantum Hawking effect for two reasons: (a) if the state of the system is Gaussian and incoherent over the incoming channels, we have proven that the GPH criterion and the quartic CS violation are equivalent to the  quadratic CS violation and (b) even if the state of the system does not belong to that class, the quadratic CS violation is still a signature of the presence of the entanglement; in particular, it is a sufficient condition for the fulfillment of the GPH criterion. Even more than that, as remarked at the end of Sec. \ref{sec:CSPH}, the quadratic CS violation is also by itself a clear indication of the quantum nature of the system, as CS inequalities are always satisfied in a classical system.

\section{Numerical results} \label{sec:Numerical}

\begin{figure*}[!tb]
\includegraphics[width=1.05\textwidth]{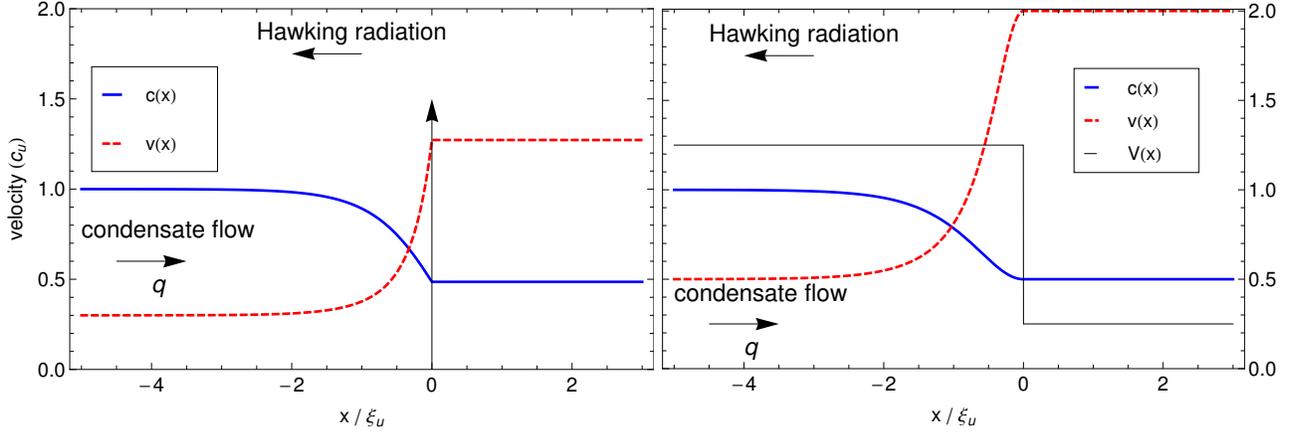}
\caption{Scheme of the delta barrier configuration (left) and the waterfall configuration (right). Here, $c(x),v(x)$ are the local sound and flow velocities, while $V(x)$ is the external potential.}
\label{fig:NonResonantScheme}
\end{figure*}

We investigate here the possibility of experimental detection of CS violation in different black-hole setups. For that purpose, we compute the measurable quantity $\tilde{\Delta}_{ud2}$, defined in Eq. (\ref{eq:Delta-ud2}). We suppose that the state of the system is given by a thermal distribution in the incoming modes, as expressed in Eq. (\ref{eq:thermaldistribution}). Since we only focus on the identification of some physical trends and not on the study of the whole parameter space, characterized by $7$ variables (see Appendix \ref{app:parametrization}), we will consider for simplicity some typical configurations studied in the literature in order to compute $\tilde{\Delta}_{ud2}$. For this analysis we distinguish between non-resonant and resonant structures. For the non-resonant case, we consider in this work two scenarios: the single delta barrier configuration and the waterfall configuration, schematically depicted in Fig. \ref{fig:NonResonantScheme}.

In the delta barrier configuration, the black hole forms near a localized potential of the form $V(x)=Z\delta(x)$ \cite{Kamchatnov2012,Zapata2011,Larre2012}; see left panel of Fig. \ref{fig:NonResonantScheme}. This configuration permits us to study theoretically the flow of a condensate through a potential barrier of finite size. On the other hand, the waterfall configuration creates the sonic horizon by using a negative step potential $V(x)=-V_0\theta(x)$, where $\theta(x)$ is the step function \cite{Larre2012}. This kind of scenario, where the black hole is produced by introducing a negative potential, has been already experimentally realized by the Technion group \cite{Lahav2010}.

\begin{figure*}[!tb]
\begin{tabular}{@{}cc@{}}
    \includegraphics[width=\columnwidth]{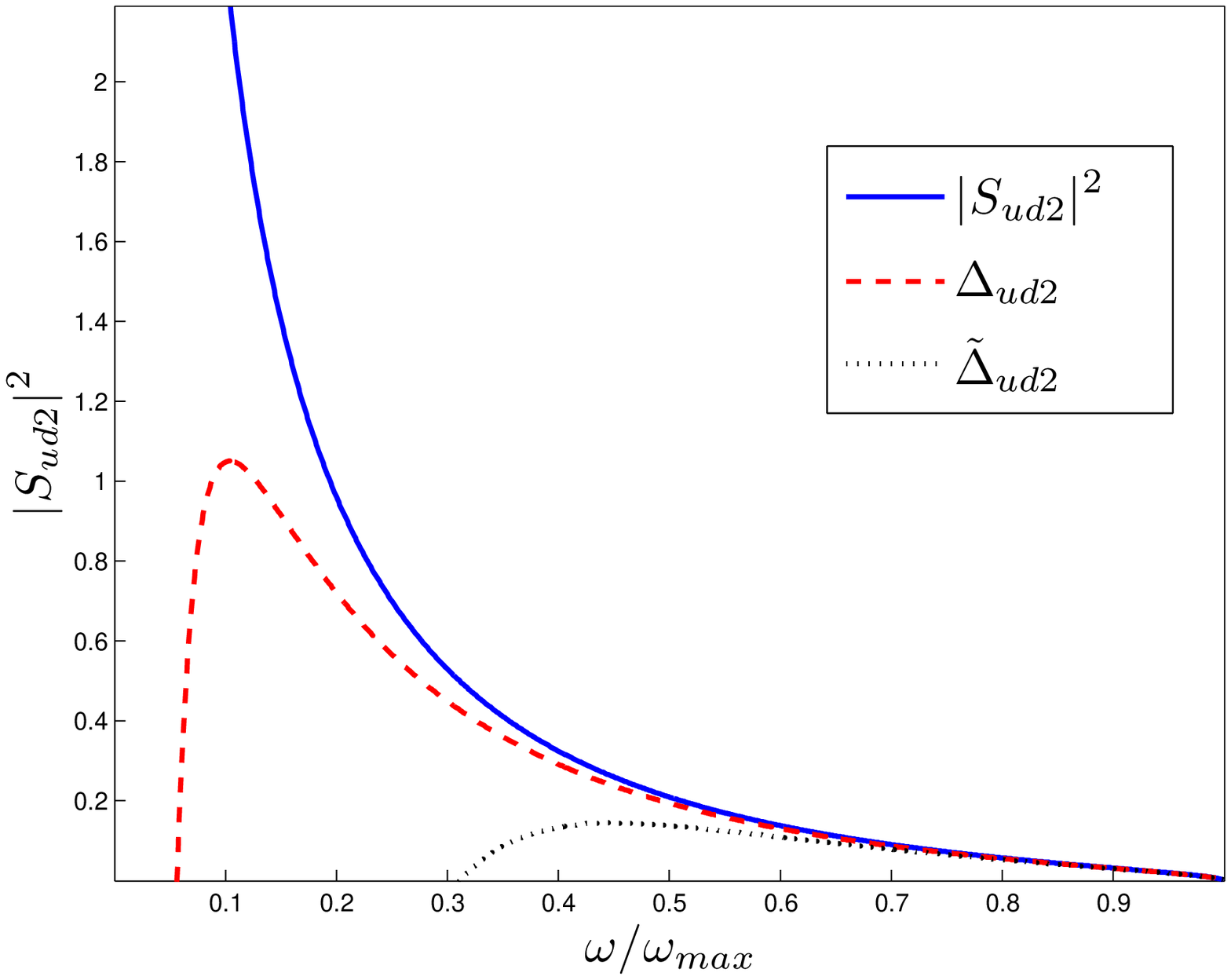} &
    \includegraphics[width=\columnwidth]{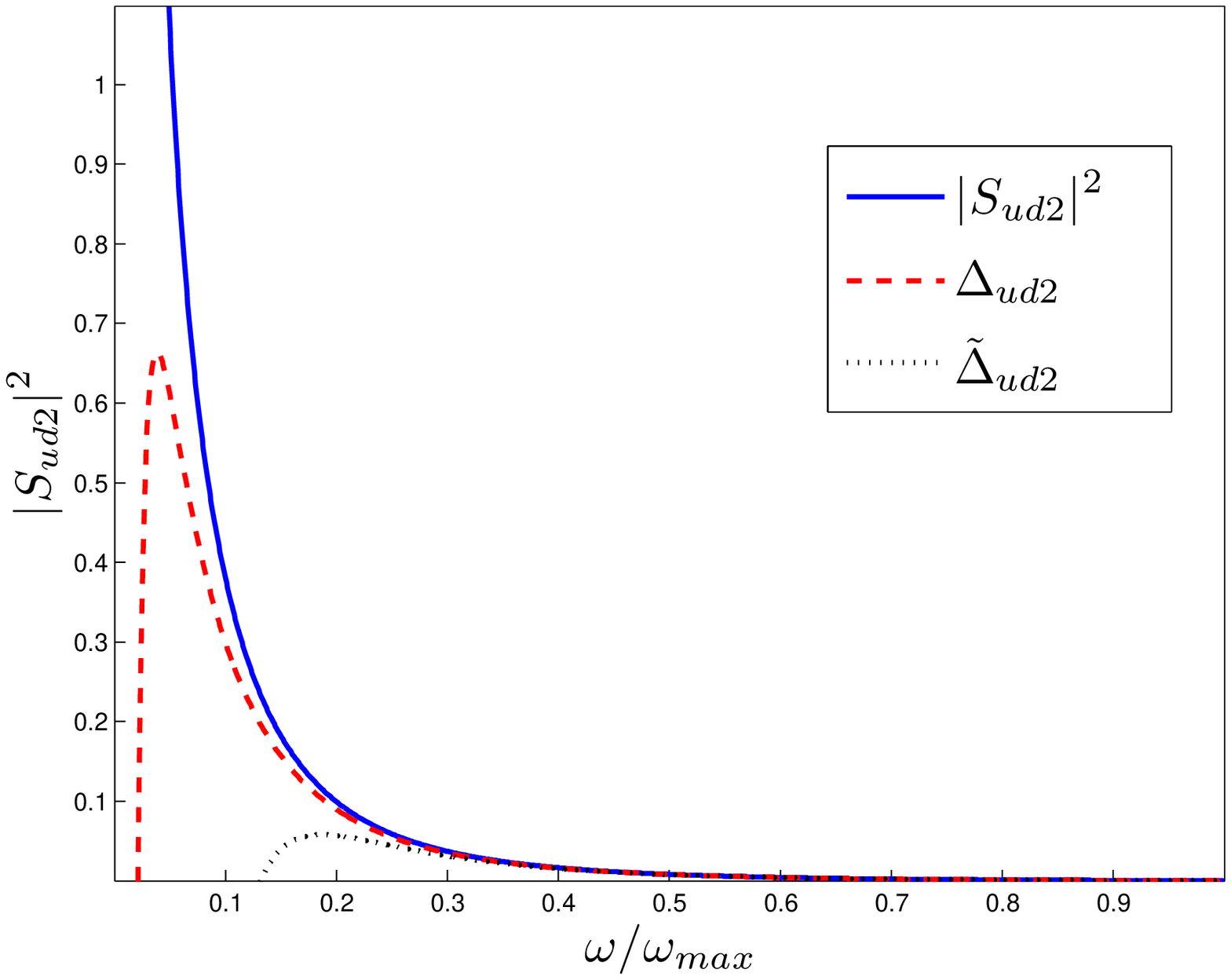} \\
\end{tabular}
\caption{Plot of the Hawking radiation spectrum and the functions $\Delta_{ud2},\tilde{\Delta}_{ud2}$ for a delta barrier configuration (left panel) and a waterfall configuration (right panel), at temperature $T=0.2$. For the delta barrier, the subsonic flow speed is $v_u=0.3$ and the barrier strength is $Z=0.62$. For the waterfall setup, the subsonic flow speed is $v_u=0.5$ and the potential depth is $V_0=1.125$. The corresponding Hawking temperatures \cite{Macher2009a,Busch2014} are $T_H=0.24$ and $T_H=0.14$, respectively}
\label{fig:NonResonantViolations}
\end{figure*}

In Fig. \ref{fig:NonResonantViolations}, we represent the Hawking spectrum, $|S_{ud2}|^2$ and the functions $\Delta_{ud2},\tilde{\Delta}_{ud2}$ for the delta barrier and the waterfall setups, both at temperature $T=0.2$, which is of the order of magnitude of typical experimental setups, where it can be as low as $T\sim0.1$ \cite{Steinhauer2014}. We see that, for some range of frequencies, the measurable function $\tilde{\Delta}_{ud2}$ satisfies $\tilde{\Delta}_{ud2}>0$, which implies the presence of CS violation. As noted previously, $\tilde{\Delta}_{ud2}>0$ is a more restrictive condition than the bare quadratic CS violation $\Delta_{ud2}>0$. However, we observe that for sufficiently large $\omega$, both $\Delta_{ud2},\tilde{\Delta}_{ud2}$ converge to $|S_{ud2}|^2$, since for high $\omega$ we have that the occupation numbers are negligible and thus, we recover the zero-temperature limit, for which $\Delta_{ud2}(T=0)=|S_{ud2}|^2$. Moreover, we see that, at high $\omega$, $|S_{d1d2}(\omega_{d1d2}(\omega))|^2=0$, because for $\omega$ such that $\omega_{d1d2}(\omega)>\omega_{\rm max}$ the anomalous scattering channel disappears, and then we have $\tilde{\Delta}_{ud2}\simeq \Delta_{ud2}(T=0)=|S_{ud2}|^2$. Finally, we note that there is no violation near $\omega=0$, as argued quite generally in Ref. \cite{deNova2014}.

\begin{figure}[htb!]
\includegraphics[width=1\columnwidth]{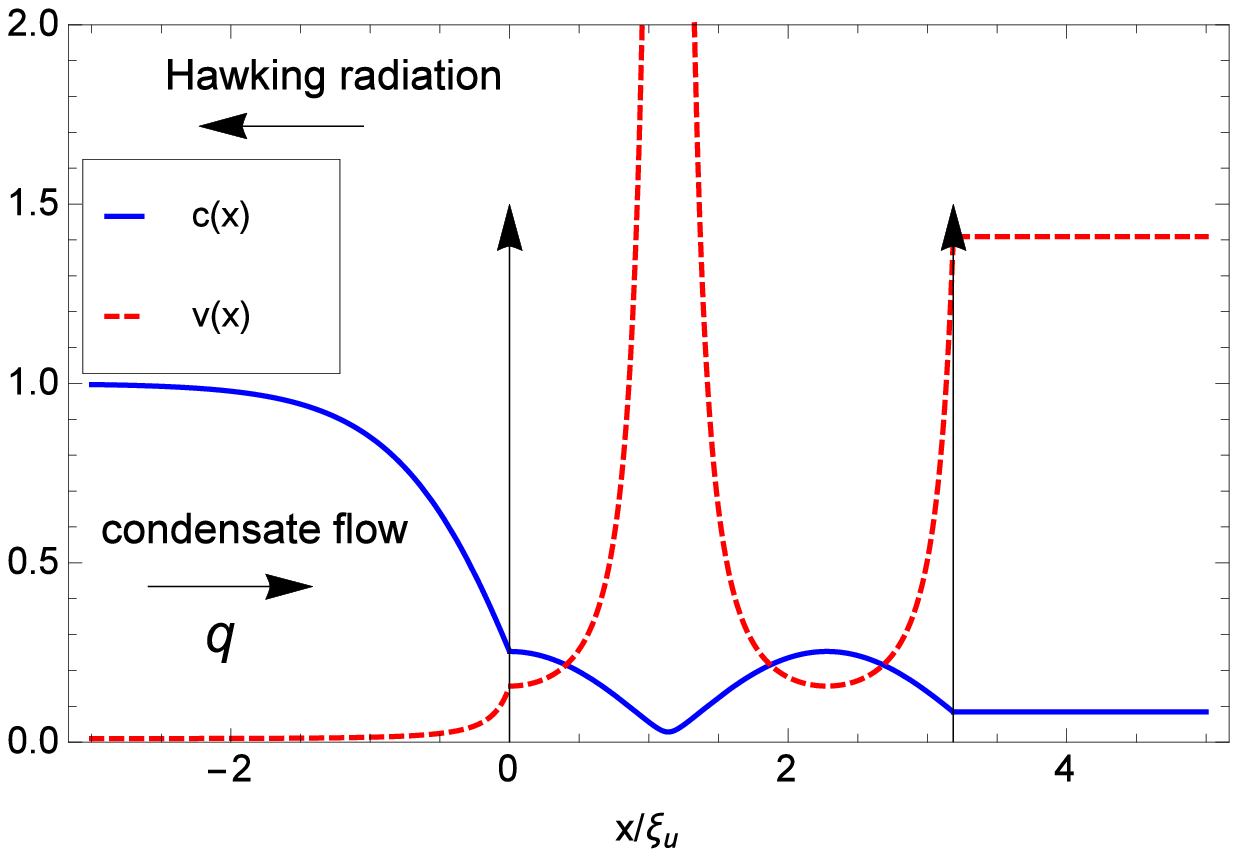}
\caption{Scheme of the double delta configuration.}
\label{fig:DoubleBarrierScheme}
\end{figure}

\begin{figure}[htb!]
\includegraphics[width=1\columnwidth]{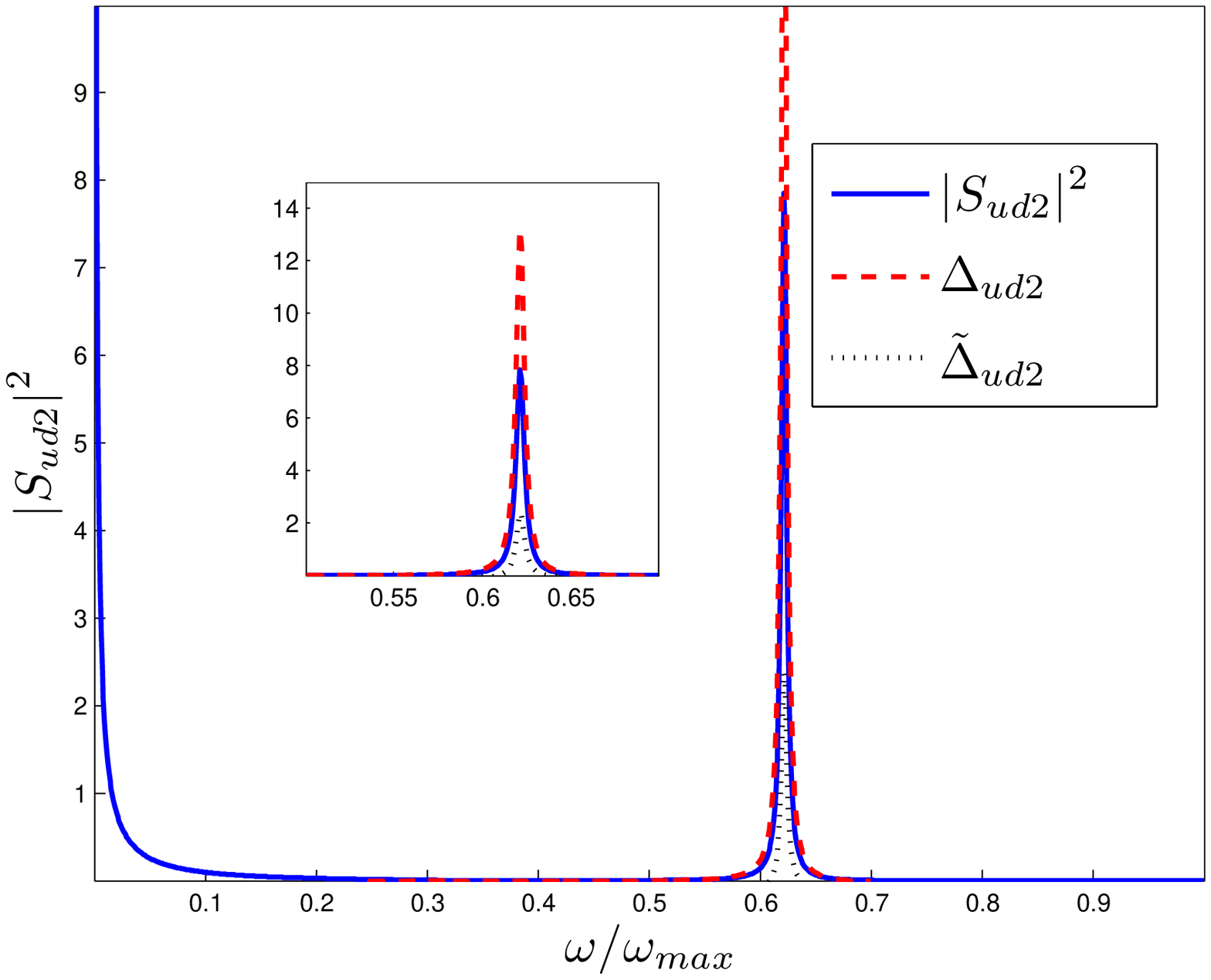}
\caption{Same as Fig. \ref{fig:NonResonantViolations}, but now for a double delta barrier configuration with parameters $Z=2.2$, $d=3.62$ and $v_u=0.01$. The temperature of the system is $T=0.7$. The inset zooms into the peak region. In this case, the Hawking temperature of the two horizons (see Fig. \ref{fig:DoubleBarrierScheme}) is $T_H=0.13$.}
\label{fig:2Deltaud2}
\end{figure}

We now focus on resonant configurations. In particular, we focus on the case of a double delta barrier, which was first considered in Ref. \cite{Zapata2011}. In this situation, the black hole is formed by two single delta barriers separated by a distance $d$, so the potential is given by $V(x)=Z[\delta(x)+\delta(x-d)]$. This setup is schematically depicted in Fig. \ref{fig:DoubleBarrierScheme}. It was shown in Ref. \cite{deNova2014} that resonant spectra can be expected to present a strong signal of CS violation. This trend can be observed in Fig. \ref{fig:2Deltaud2}, where we represent the same magnitudes as in Fig. \ref{fig:NonResonantViolations} but now for a higher temperature $T=0.7$. We see that, even for this relatively high temperature, the experimental signal $\tilde{\Delta}_{ud2}$ is, at the resonance frequency, substantially larger than in the non-resonant case.

\section{Conclusions} \label{sec:conclusions}

We have analyzed the range of validity of different existent criteria for the identification of deep quantum behavior as applied to the outgoing channels involved in the spontaneous Hawking radiation. Specifically, we have compared the generalized Peres-Horodecki criterion with the violation of quadratic and quartic Cauchy-Schwarz inequalities. We have shown that, under certain physical conditions (Gaussian processes and, simultaneously, incoherent incoming channels), all the considered criteria are equivalent. When such conditions are not fulfilled, we have shown that the quartic CS violation and the non-separability of the state represent independent mathematical conditions.

We have also investigated the possible measurement of the different criteria in realistic scenarios. By taking into account the different spatial location of the subsonic and supersonic regions, we have shown that only certain correlation functions can be obtained. For simplicity we have focused only on one detection scheme, but we expect similar problems to arise in other kind of measurements, since in any realistic situation the supersonic and subsonic regions are necessarily placed in different regions. However, our work also shows that this limitation is not a major problem, as we can often measure the quadratic CS violation, which is also a sufficient condition for the entanglement of the system. Finally, our numerical results of Section \ref{sec:Numerical} show that, in typical analog configurations, the CS violation can be detected in an achievable range of temperatures.

Sonic black holes, as well as related setups such as black-hole lasers, have the potential to become an important element of atomtronic circuits, for instance, as a source of entangled quasiparticles in the broader context of quantum communication.

\acknowledgments
We thank A. Amo, F. Michel, R. Parentani, D. Gu\'ery-Odelin and J. Steinhauer for valuable discussions. This work has been supported by MINECO (Spain) through grants FIS2010-21372 and FIS2013-41716-P, and by
Comunidad de Madrid through grant MICROSERES-CM (S2009/TIC-1476).

\appendix

\section{General remarks on Cauchy-Schwarz inequalities and entanglement}\label{app:CSnonseparability}

\subsection{Cauchy-Schwarz inequality for operators}

We briefly review the mathematical CS inequalities for operators in a Hilbert space, which are always fulfilled, see Ref. \cite{Adamek2013} for example.  We start by defining the scalar product associated to a state $\hat{\rho}$ for two operator $\hat{A},\hat{B}$  as:

\begin{equation}\label{eq:scalarproduct}
(\hat{A},\hat{B})\equiv\braket{\hat{A}^{\dagger}\hat{B}}=\text{Tr}(\hat{\rho}\hat{A}^{\dagger}\hat{B}).
\end{equation}

It is easy to check that the previous product satisfies the usual properties of a scalar product. Thus, it satisfies the mathematical Cauchy-Schwarz inequality:
\begin{equation}\label{eq:CSmathematical}
|\braket{\hat{A}^{\dagger}\hat{B}}|^2\leq \braket{\hat{A}^{\dagger}\hat{A}}\braket{\hat{B}^{\dagger}\hat{B}} \, .
\end{equation}
In particular, the previous inequality implies for $B=1$:
\begin{equation}\label{eq:meansquare}
|\braket{\hat{A}}|^2=|\braket{\hat{A}^{\dagger}}|^2\leq \braket{\hat{A}^{\dagger}\hat{A}} \, ,
\end{equation}
and, for $A=\hat{\gamma}_{i}$ and $B=\hat{\gamma}_{j}$,
\begin{equation}\label{eq:CSgijimpossible}
|g_{ij}|^2=|\braket{\hat{\gamma}^{\dagger}_{i}\hat{\gamma}_{j}}|^2
\leq\braket{\hat{\gamma}^{\dagger}_{i}\hat{\gamma}_{i}}\braket{\hat{\gamma}^{\dagger}_{j}\hat{\gamma}_{j}}=g_{ii}g_{jj} \, .
\end{equation}
Thus, the CS inequality $|g_{ij}|^2\leq g_{ii}g_{jj}$ is always satisfied and can never be violated. On the other hand, taking $A=\hat{\gamma}^{\dagger}_{i}$ and $B=\hat{\gamma}_{j}$, we arrive at:
\begin{equation}\label{eq:CScijpos}
|c_{ij}|^2=|\braket{\hat{\gamma}_{i}\hat{\gamma}_{j}}|^2
\leq\braket{\hat{\gamma}_{i}\hat{\gamma}^{\dagger}_{i}}
\braket{\hat{\gamma}^{\dagger}_{j}\hat{\gamma}_{j}}=(g_{ii}+1)g_{jj} \, .
\end{equation}
Thus, the mathematical CS inequality states $|c_{ij}|^2\leq(g_{ii}+1)g_{jj}$, leaving the possibility of quadratic CS violation, Eq. (\ref{eq:CSviolation2}). The same type of argument holds for quartic CS violations, where an analogous reasoning leads to:
\begin{eqnarray}\label{eq:CSGammaijpos}
\nonumber |\Gamma_{ij}|^2&=&|\braket{\hat{\gamma}^{\dagger}_{i}\hat{\gamma}_{i}\hat{\gamma}^{\dagger}_{j}\hat{\gamma}_{j}}|^2
\leq\braket{\hat{\gamma}^{\dagger}_{i}\hat{\gamma}_{i}\hat{\gamma}^{\dagger}_{i}\hat{\gamma}_{i}}\braket{\hat{\gamma}^{\dagger}_{j}\hat{\gamma}_{j}\hat{\gamma}^{\dagger}_{j}\hat{\gamma}_{j}}\\
&=&(\Gamma_{ii}+g_{ii})(\Gamma_{jj}+g_{jj})
\end{eqnarray}
which also leaves the possibility of quartic CS violations; see Eq. (\ref{eq:CSviolation4}).

\subsection{Quadratic Cauchy-Schwarz violation and the generalized Peres-Horodecki criterion}

We now consider bipartite states of two modes $i,j$. We define the vector $\hat{X}\equiv [\hat{q}_i,\hat{p}_i,\hat{q}_j,\hat{p}_j]^{T}$, where the phase space operators $\hat{q}_k,\hat{p}_k$ are related to the annihilation operator of the modes $k=i,j$ through $\hat{\gamma}_k=(\hat{q}_k+i\hat{p}_k)/\sqrt{2}$. Following Ref. \cite{Simon2000}, we define the $4\times 4$ matrix:

\begin{eqnarray}\label{eq:Uncer}
M&\equiv& V+i\frac{W}{2}\\
\nonumber W&=&\left[\begin{array}{cc} J & 0\\
0 & J \end{array}\right],~J=\left[\begin{array}{cc}
0 & 1\\
-1 & 0
\end{array}\right]\,
\end{eqnarray}
where $V$ is the covariance matrix with matrix elements $V_{\alpha\beta}=\frac{1}{2}\{\Delta\hat{X}_{\alpha},\Delta\hat{X}_{\beta}\}$, with $\Delta \hat{X}\equiv \hat{X}-\braket{\hat{X}}$ and $\{\Delta\hat{X}_{\alpha},\Delta\hat{X}_{\beta}\}$ the anticommutator. The matrix $J$ is the simplectic matrix in two dimensions. The matrix $M$ is non-negative, $M\geq 0$, as this fact represents an alternative expression of the uncertainty principle \cite{Simon2000}.

If the state $\hat{\rho}$ is separable, it has the form (\ref{eq:separability}), and by taking partial transpose of the state with respect to the subsystem $j$, we obtain an operator $\hat{\rho}_{t}$ which is also necessarily a physical state. The GPH criterion is based on this fact. It can be shown that if $\hat{\rho}_{t}$ is a physical state, the matrix $M_t$ is also non-negative, where $M_t$ is given by \cite{Simon2000}:

\begin{eqnarray}\label{eq:Uncertrans}
M_t&\equiv& V_t+i\frac{W}{2}\\
\nonumber V_t&=&\Lambda V \Lambda,~\Lambda\equiv \rm{diag}[1,1,1,-1].
\end{eqnarray}
The condition $M_t\geq 0$ is the uncertainty principle for the state $\hat{\rho}_{t}$. As $M$ is non-negative, $M_t$ is non-negative if and only if $\det M_t\geq 0$. The conditions $\det M,M_t\geq 0$ are equivalent to $\mathcal{P}^{\pm}_{ij}\geq0$, respectively, where:

\begin{eqnarray}\label{eq:GPHpm}
\mathcal{P}^{\pm}_{ij}&\equiv&\det A_i\det A_j+\left(\frac{1}{4}\mp \det C_{ij}\right)^2\\
\nonumber&-&\text{tr}(JA_iJC_{ij}JA_jJC_{ij}^{T})-\frac{1}{4}(\det A_i+\det A_j) \, .
\end{eqnarray}
The matrices $A_i,A_j,C_{ij}$ are $2\times 2$ submatrices of the covariance matrix $V$:
\begin{equation}\label{eq:WBlocks}
V=\left[\begin{array}{cc} A_{i} & C_{ij}\\
C^{T}_{ij} & A_{j} \end{array}\right]
\end{equation}
We note that $\mathcal{P}^{+}_{ij}\geq0$ is always satisfied since $\hat{\rho}$ is a physical state. However, when $\mathcal{P}^{-}_{ij}<0$, the state $\hat{\rho}_{t}$ is not a physical state, which implies that the original state $\hat{\rho}$ is not separable. We can put together both conditions by defining the GPH function $\mathcal{P}_{ij}$ as:
\begin{eqnarray}\label{eq:GPH}
\mathcal{P}_{ij}&\equiv&\det A_i\det A_j+\left(\frac{1}{4}-|\det C_{ij}|\right)^2\\
\nonumber&-&\text{tr}(JA_iJC_{ij}JA_jJC_{ij}^{T})-\frac{1}{4}(\det A_i+\det A_j) ,
\end{eqnarray}
Thus, if $\mathcal{P}_{ij}<0$, the state is entangled. This result is the GPH criterion. However, when $\det C_{ij}\geq 0$, the state is separable, as $\mathcal{P}_{ij}=\mathcal{P}^{+}_{ij}\geq0$, so only states with $\det C_{ij}< 0$ can be entangled.

On the basis of the previous results, we now prove that the quadratic CS violation implies the fulfillment of the GPH criterion. Suppose that $\mathcal{P}_{ij}\geq0$. In that case, the matrix $M_t$ is non-negative and we can define an associated scalar product for vectors $u,v\in \mathbb{C}^4$ as:
\begin{equation}\label{eq:scalarproduct4}
(u,v)_t\equiv u^{\dagger}M_tv
\end{equation}
which satisfies the associated CS inequality:
\begin{equation}\label{eq:CSscalar4}
|(u,v)_t|^2\leq (u,u)_t(v,v)_t \, .
\end{equation}
Assuming the usual case where the annihilation and destruction operators have zero expectation values, the matrices $A_{i,j}$ and $C_{ij}$ can be written in terms of the correlation functions of Eq. (\ref{eq:gdef}) as:
\begin{eqnarray}\label{eq:GPHMatrices}
A_{k}&=&\left(g_{kk}+\frac{1}{2}\right)\mathbb{I}_2+\left[\begin{array}{cc}
\text{Re}~c_{kk} & \text{Im}~c_{kk}\\
\text{Im}~c_{kk} & -\text{Re}~c_{kk}
\end{array}\right]\\
C_{ij}&=&
\left[\begin{array}{cc}
\text{Re}(g_{ij}-c_{ij}) & \text{Im}(g_{ij}+c_{ij})\\
\text{Im}(-g_{ij}+c_{ij}) & \text{Re}(g_{ij}-c_{ij})
\end{array}\right]\, ,
\end{eqnarray}
with the index $k=i,j$ and $\mathbb{I}_2$ the $2\times 2$ identity matrix. 
By inserting
\begin{equation}\label{eq:CSvectors4}
u=\frac{1}{\sqrt{2}}\left[\begin{array}{c}
0\\ 0\\ 1\\ i
\end{array}\right],~v=\frac{1}{\sqrt{2}}\left[\begin{array}{c}
1\\ i\\ 0\\ 0
\end{array}\right]
\end{equation}
into Eq. (\ref{eq:CSscalar4}), we obtain the quadratic CS inequality $|c_{ij}|^2\leq g_{ii}g_{jj}$. Thus, if there is quadratic CS violation, the matrix $M_t$ cannot be non-negative, which implies that $\mathcal{P}_{ij}<0$ and the GPH criterion is satisfied. We conclude that the quadratic CS violation is a sufficient condition for the fulfillment of the GPH criterion.

\subsection{Quartic Cauchy-Schwarz violation and entanglement}

The previous arguments cannot be applied to the quartic CS violation Eq. (\ref{eq:CSviolation4}). As a counterexample, the direct product of two pure number states of the modes $i,j$, $\hat{\rho}=\ket{n}\bra{n}\otimes\ket{n'}\bra{n'}$, which is a manifestly separable state, violates the quartic CS inequality. Even for Gaussian states, the two conditions, quartic CS violation and entanglement, are still independent. For instance, for a general Gaussian state, the quartic CS violation of Eq. (\ref{eq:CSviolation4}) can be expressed, via Wick's theorem, in terms of the quadratic correlations:

\begin{equation}\label{eq:CSviolation4Wick}
\Theta_{ij}=|c_{ij}|^2+|g_{ij}|^2+g_{ii}g_{jj}-\sqrt{2g_{ii}^2+|c_{ii}|^2}\sqrt{2g_{jj}^2+|c_{jj}|^2}
\end{equation}

For Gaussian states the GPH condition is equivalent to the non-separability of the system. Using Eqs. (\ref{eq:GPH}), (\ref{eq:GPHMatrices}), we can compute explicitly the GPH function:
\begin{widetext}
\begin{eqnarray}\label{eq:GPHcomplete}
\nonumber \mathcal{P}_{ij}&=&[g_{ii}g_{jj}-|c_{ij}|^2][(g_{ii}+1)(g_{jj}+1)-|c_{ij}|^2]+4\left(g_{ii}+\frac{1}{2}\right)\text{Re}(g_{ij}c^{*}_{jj}c_{ij})+4\left(g_{jj}+\frac{1}{2}\right)\text{Re}(g_{ij}c_{ii}c^{*}_{ij})\\
&-&2\text{Re}(c^{2}_{ij}c^{*}_{ii}c^{*}_{jj})-2\text{Re}(g^{2}_{ij}c_{ii}c^{*}_{jj})+|c_{ii}|^2|c_{jj}|^2+|g_{ij}|^4-|g_{ij}|^2(g_{ii}+g_{jj}+2g_{ii}g_{jj}+2|c_{ij}|^2)\\
\nonumber&-&g_{ii}(g_{ii}+1)|c_{jj}|^2-g_{jj}(g_{jj}+1)|c_{ii}|^2 \, .
\end{eqnarray}
\end{widetext}
This expression applies whenever $|c_{ij}|\geq|g_{ij}|$. For $|c_{ij}|<|g_{ij}|$, we have $\det C_{ij}>0$ and the state is separable, as previously explained.

As we can see, the expressions for the quartic CS violation and the GPH criterion represent different conditions for arbitrary Gaussian states. In particular, it is easy to find states violating the quartic CS inequality with $\det C_{ij}>0$, which means that they are separable. Also, we can find entangled states that satisfy the quartic CS inequality.

\section{Parametrization of outgoing correlations for incoherent incoming Gaussian states}\label{app:parametrization}

We discuss in this Appendix the parameters needed to compute the correlation functions defined in the main text. A related discussion appeared in Ref. \cite{Busch2014}. The scattering matrix $S$ relates the ``out'' scattering states with the ``in'' scattering states through Eq. (\ref{eq:inoutmodesrelation}). The conservation of the commutation relations implies the relation:
\begin{equation}\label{eq:pseudounitarity}
S^{\dagger}\eta S=\eta\equiv{\rm diag}(1,1,-1)\, ,
\end{equation}
which means that $S\in U(2,1)$. It follows that $S^{-1}=\eta S^{\dagger} \eta$ and, using standard linear algebra, we can prove:
\begin{equation}
S^*_{ij}=\frac{\eta_{ii} m_{ij} \eta_{jj}}{\textrm{det}~S} \, ,
\end{equation}
where $m_{ij}$ is the minor associated to the $S$ matrix element $S_{ij}$. In order to simplify the notation, in this section we relabel the indices $u,d1,d2$ as $1,2,3$ in order to match the matrix indices ordering. With the help of the previous results, it can be proven that any matrix $S \in U(2,1)$ satisfies:
\begin{eqnarray}\label{eq:param}
BSA=S_{\rm PA}=&\left[\begin{array}{ccc} \det S &0&0\\
0&S^*_{33}&N\\
0&N&S_{33}\end{array}\right]&\\
\nonumber B\equiv \frac{1}{N}&\left[\begin{array}{ccc} -S_{23}&S_{13}&0\\
S^*_{13}&S^*_{23}&0\\
0&0&N\end{array}\right]&\\
\nonumber A\equiv \frac{1}{N}&\left[\begin{array}{ccc} -S_{32}&S^*_{31}&0\\
S_{31}&S^*_{32}&0\\
0&0&N\end{array}\right]&
\end{eqnarray}
where $N=\sqrt{|S_{13}|^2+|S_{23}|^2}=\sqrt{|S_{31}|^2+|S_{32}|^2}=\sqrt{|S_{33}|^2-1}$. We see that the matrices $A,B$ are unitary. Inverting the previous relation, we obtain:
\begin{equation}\label{eq:Sparam}
S=B^{\dagger}S_{\rm PA}A^{\dagger} \, .
\end{equation}

Equations (\ref{eq:param}),(\ref{eq:Sparam}) show explicitly that we only need $9$ real parameters to parametrize the $S$ matrix: $|S_{31}|,|S_{13}|,|S_{33}|,\chi,\phi_{13},\phi_{23},\phi_{31},\phi_{32},\phi_{33}$, with $\phi_{ij}$ the phase of $S_{ij}$, $S_{ij}=|S_{ij}|e^{i\phi_{ij}}$ and $\chi$ that of $\det S$, $\det S=e^{i\chi}$. That is, we use $3$ amplitudes and $6$ phases to characterize the complete scattering matrix. Interestingly, the matrix $S_{\rm PA}$ shows clearly that the Hawking radiation acts as a non-degenerate parametric amplifier \cite{Walls2008}.

A convenient way to write the $S$ matrix for our problem is:
\begin{eqnarray}\label{eq:inoutphases}
S&=&U^{\dagger}_{\rm{out}}\tilde{S}U_{\rm{in}}\\
\nonumber U_{k}&=&{\rm diag}[e^{i\delta_{1-k}},e^{i\delta_{2-k}},e^{i\delta_{3-k}}]
\end{eqnarray}
(with $k=\rm{in,out}$),
since we can fix the elements of $U_{\rm{in,out}}$ in such a way that they absorb the phases $\phi_{31},\phi_{32},\phi_{13},\phi_{23},\phi_{33}$, and so $\tilde{S}_{31},\tilde{S}_{32},\tilde{S}_{13},\tilde{S}_{23},\tilde{S}_{33}$ become purely real. This means that the matrix $\tilde{S}$ is characterized by only $4$ parameters, $|S_{31}|,|S_{13}|,|S_{33}|, \tilde{\chi}$, with
\begin{equation}
\det \tilde{S}=e^{i\tilde{\chi}}=\det S~e^{-i(\phi_{13}+\phi_{23}+\phi_{13}+\phi_{23}-\phi_{33})}.
\end{equation}

Nevertheless, we do not need all $9$ parameters for computing the physical quantities of the main text. For instance, the CS violations of Eq. (\ref{eq:CSGamma}) are invariant under phase transformations of the ``out'' states. It can be also shown that the GPH function is invariant under these transformations, see Eq. (\ref{eq:GPHcomplete}). As we are always considering incoherent incoming modes, all the quantities are also invariant under phase transformations of the incoming modes. From Eq. (\ref{eq:inoutphases}), and taking into account the previous observations, we conclude that all the requested quantities depend only on the matrix $\tilde{S}$, which is characterized by just $4$ parameters, as noted in the previous paragraph. Using Eqs. (\ref{eq:pseudounitarity}), (\ref{eq:Sparam}) we can write $\tilde{S}$ as:
\begin{widetext}
\begin{equation}\label{eq:definitiveparametrization}
\tilde{S}=\left[\begin{array}{ccc} -\sin\alpha&\cos\alpha&0\\
\cos\alpha&\sin\alpha&0\\
0&0&1\end{array}\right]\left[\begin{array}{ccc} e^{i\tilde{\chi}} &0&0\\
0&\cosh \gamma&\sinh \gamma\\
0&\sinh \gamma&\cosh \gamma\end{array}\right]\left[\begin{array}{ccc} -\sin\beta&\cos\beta&0\\
\cos\beta&\sin\beta&0\\
0&0&1\end{array}\right]
\end{equation}
\end{widetext}
with $|S_{33}|=\cosh \gamma, |S_{13}|=\sinh \gamma \cos\alpha$, and $|S_{31}|=\sinh \gamma \cos\beta$. Thus, we only need $4$ parameters in order to parametrize the scattering matrix for the calculations of this work: $\alpha,\beta,\gamma, \tilde{\chi}$.  We note that this parametrization differs slightly from that used in Ref. \cite{Busch2014}, where $4$ amplitudes were used to characterize the state of the system. Nevertheless, that election is equivalent to that used here.

Finally, taking into account that the state of the system $\hat{\rho}$ is characterized by only $3$ numbers, $n_i$ [with $i=1,2,3$, see Eq. (\ref{eq:open-bc})], we have that the whole problem is completely determined by $7$ parameters, $4$ arising from the $S$ matrix and $3$ arising from the specification of the incoherent incoming Gaussian state.

\bibliographystyle{apsrev}
\bibliography{PhDJR}

\end{document}